\documentclass{pas}

\usepackage{amsmath}
\usepackage{multirow}

\providecommand{\tablefoot}[1]{\\[2pt]\noindent\footnotesize\textit{Notes.} #1}
\providecommand{\checkmark}{\ensuremath{\surd}}

\begin{document}


\lefttitle{Reversed illumination dependence of UEMR from Starlink DTC}

\righttitle{H. Dong et al.}


\jnlPage{1}{12}
\jnlDoiYr{2026}
\doival{10.1017/pasa.xxxx.xx}

\articletitt{Research Paper}


\title{A reversed solar illumination dependence of unintended emission from Starlink Direct-to-Cell satellites at 72--234 MHz with the EDA2}

\author{Haofan Dong$^{1}$ and Houtianfu Wang$^{1}$
        and Hanlin Cai$^{1}$ and Ozgur B. Akan$^{1,2}$}

\affil{$^1$Centre for neXt Communications (CXC), Department of
       Engineering, University of Cambridge, Cambridge CB3 0FA,
       United Kingdom and $^2$Centre for neXt Communications (CXC),
       Department of Electrical and Electronics Engineering,
       Ko\c{c} University, 34450 Istanbul, T\"urkiye}

\corresp{Haofan Dong, Email: hd489@cam.ac.uk}

\history{(Received xx xx xxxx; revised xx xx xxxx; accepted xx xx xxxx)}


\begin{abstract}
Second-generation Starlink Direct-to-Cell (DTC) satellites carry an additional payload for direct cellular phone connectivity whose unintended electromagnetic radiation (UEMR) at sub-$300$\,MHz frequencies has not been individually characterised. We reanalyse $112\,534$ detections from $1\,806$ Starlink satellites observed with the Engineering Development Array version~2 (EDA2) at $21$ frequencies between $72.685$ and $234.375$\,MHz \citep{Grigg2025}, separating $175$ DTC and $1\,623$ Ku-only v2-Mini comparison satellites via the McDowell General Catalogue \citep{McDowell2020}. DTC satellites emit a range-corrected flux density $1.45\times$ that of the Ku-only comparison (Cliff's $\delta=+0.30$, $p=2.6\times10^{-11}$). At $230.469$\,MHz the XX detection fraction reaches $0.811$ against a $0.481$ baseline ($p\sim10^{-274}$), and $11$ of $21$ frequency channels show Benjamini--Hochberg-significant polarisation anomalies. The DTC population is brighter in eclipse than in sunlight (illuminated/eclipsed flux density ratio $0.47$) while the Ku-only comparison shows the opposite sense ($1.18$); the reversal persists across altitude, sub-satellite latitude, frequency, and launch-epoch matching. The reversal strongly disfavours UEMR mechanisms that scale monotonically with instantaneous solar photocurrent and favours an active on-board source whose effective duty cycle is larger at lower equilibrium temperature. Within the $230.469$\,MHz coarse channel, fine-channel inspection isolates the excess to a single $\sim 24$\,kHz bin near $230.627$\,MHz, tail-driven and absent at five control channels. Three falsifiable mechanism-discrimination tests show this feature is not coincident with the LOFAR-resolved \citet{Bassa2024} clock fundamentals, is unresolved at the EDA2 $24$\,kHz resolution, and is heterogeneously expressed across the v2-Mini fleet rather than driven by a few permanently bright units or by uniform thermal scaling.
\end{abstract}


\maketitle

\section{Introduction}
\label{sec:intro}

The rapid expansion of low Earth orbit (LEO) satellite constellations
is changing the radio sky at frequencies that are central to
ground-based low-frequency radio astronomy. As of early 2026, the McDowell General Catalogue of Artificial
Space Objects records more than $8\,000$ operational Starlink
satellites in low Earth orbit, with several other large
constellations also entering deployment
\citep{McDowell2020,Lawler2022,Grigg2025}. The impact of these
megaconstellations on optical and radio astronomy has been the
subject of sustained community concern
\citep{Walker2020_BAAS,Walker2020_DQS,Barentine2023}.
Even when a satellite does not intentionally transmit in a protected
radio astronomy band, its on-board electronics, high-speed digital
interfaces, and power-conditioning systems can produce wideband
unintended electromagnetic radiation (UEMR) that propagates into the
receiver \citep{DiVruno2023}. For the Square Kilometre Array
low-frequency telescope \citep[SKA-Low;][]{Dewdney2009}, whose core
stations operate between $50$ and $350$\,MHz and target key
cosmological signals from the Cosmic Dawn and Epoch of Reionisation
\citep{Furlanetto2006,Koopmans2015}, UEMR has emerged as a limiting
contribution to the effective noise floor at the array site
\citep{Grigg2025,Bassa2024}, and even comparatively faint
interference is known to inject excess power into $21$\,cm power
spectrum measurements \citep{Wilensky2020}.

Following \citet{DiVruno2023}, we distinguish UEMR from intended
emission and from regulated unwanted emissions. UEMR is neither
governed by radio regulations such as ITU-R Recommendation RA.769
\citep{ITU-R-RA769} nor by the framework for unwanted emissions
from non-geostationary fixed-satellite systems \citep{ITU-R_S1586},
nor designed into the system by the operator; it
is an unavoidable side product of on-board hardware. In contrast to
intended transmissions, the UEMR spectrum is typically broadband, its
time and frequency structure reflects internal digital clock harmonics
and switched-mode power-supply fundamentals, and its amplitude can
depend on the operational state of the satellite, which is not
directly observable from the ground.

Systematic detection of UEMR from modern LEO constellations dates
from \citet{DiVruno2023}, who used the Low-Frequency Array
\citep[LOFAR;][]{vanHaarlem2013} to identify narrow spectral features
from $47$ Starlink satellites in the $110$--$188$\,MHz range.
\citet{Grigg2023}, building on the passive-radar satellite
detection technique developed for the Murchison Widefield Array by
\citet{Prabu2020}, independently detected broadband Starlink UEMR
with the EDA2 in Western Australia, in a location that is
representative of the SKA-Low site. \citet{Bassa2024} subsequently reported that the
second-generation v2-Mini satellites emit UEMR up to a factor of $32$
stronger than their v1.0 and v1.5 predecessors, marking a regression
of the on-board UEMR environment across a single vendor's product
cycle. \citet{Zhang2025} used the New Extension in Nan\c{c}ay Upgrading
LOFAR \citep[NenuFAR;][]{Zarka2012,Zarka2020} to add polarimetric
constraints on the Stokes parameters of the v2-Mini broadband signal.
Most recently, \citet{Grigg2025} released a survey of $112\,534$
stacked EDA2 detections across $1\,806$ satellites and $21$ frequency
channels, together with the underlying event catalogue, providing the
first statistically large sample in which individual satellites can
be tracked across orbits.

During the period covered by the \citet{Grigg2025} observations,
SpaceX began deploying a distinct subclass of v2-Mini satellites
carrying the Direct-to-Cell (DTC) payload, identified in the McDowell
General Catalogue of Artificial Space Objects
\citep[GCAT;][]{McDowell2020} by the satellite bus label
\texttt{V2MD}. These satellites differ from the Ku-only v2-Mini
(\texttt{V2M}) primarily in that they carry additional L-band and
S-band equipment for direct cellular phone connectivity under the
Supplemental Coverage from Space authorisation granted to the
operator \citep{FCC_SCS_2024}. Although DTC
satellites are included in the aggregate statistics of
\citet{Grigg2025}, to our knowledge no published study has yet
isolated their UEMR
contribution, tested whether the DTC payload changes the low-frequency
UEMR spectrum, or examined whether their emission properties depend
on the spacecraft's illumination state. These questions have immediate
practical relevance, because the DTC fleet is growing rapidly and will
form a substantial fraction of the operational constellation within
the next two years.

In this paper we present a reanalysis of the \citet{Grigg2025} event
catalogue in which DTC satellites are separated from Ku-only v2-Mini
comparison satellites using the GCAT bus classification. We examine
three questions: whether the DTC population shows a per-satellite
excess in range-corrected flux density relative to the Ku-only
comparison; whether any coarse frequency channel exhibits a
polarisation asymmetry beyond the instrumental baseline; and whether
the flux density depends on the solar illumination state of the
spacecraft. We find that DTC emission is, on average, brighter when
the spacecraft is eclipsed than when it is illuminated, whereas the
Ku-only comparison population shows the opposite sense, and that this
reversal persists under stratification by altitude, sub-satellite
latitude, and frequency. We then discuss the implications for the
physical origin of the UEMR and for future SKA-Low operations.

\section{Observations and analysis}
\label{sec:obs}

\subsection{Data source}
\label{sec:data}

We use the event catalogue released by \citet{Grigg2025}, which
reports $112\,534$ stacked detections\footnote{``Stacked'' refers
to the fine-channel index value $31$ in the \citet{Grigg2025}
catalogue; the individual fine channels $0$--$30$ within each coarse
channel are retained in the public release, and are used here for
the dynamic spectrum analysis presented in
Sect.~\ref{sec:results}.} of Starlink satellites across $21$
discrete EDA2 frequency channels covering $72.685$--$234.375$\,MHz.
The observations were carried out with the Engineering Development
Array version~2 \citep{Wayth2021}, located at the Murchison
Radio-astronomy Observatory (MRO) in Western Australia, between UTC
2024~June and 2024~October. The catalogue is retrieved from Zenodo
(\href{https://doi.org/10.5281/zenodo.15089853}{10.5281/zenodo.15089853})
and its tabulated positions, fluxes, and ephemerides are used as
provided.

\subsection{Satellite classification}
\label{sec:classify}

Individual satellites are classified using the McDowell General
Catalogue of Artificial Space Objects \citep[GCAT;][]{McDowell2020},
retrieved on 2026~April~8. The NORAD identifiers reported by
\citet{Grigg2025} are cross-matched against the GCAT \texttt{Bus}
field, and three populations are retained:
(i) \textit{v2-Mini DTC} satellites carrying the Direct-to-Cell
payload (bus label \texttt{V2MD}), comprising $175$ objects with
$10\,180$ stacked detections;
(ii) \textit{v2-Mini Ku pre-Optical} satellites (\texttt{V2M}),
comprising $1\,623$ objects with $102\,197$ detections; and
(iii) \textit{v1.x} satellites (\texttt{V1.0} and \texttt{V1.5}),
of which $8$ objects with $157$ detections are present.
The v2-Mini Optical variant (\texttt{V2MO}) is absent from the
catalogue, consistent with its deployment timing. The GCAT bus
classification is derived from FCC filings and launch manifests
published by the operator, and has been cross-checked against
independent launch-tracking resources; its labels are adopted
without further modification. The $175$ Direct-to-Cell satellites in the
sample span multiple launch batches, orbital planes, and vehicle
serial numbers, which together provide sufficient independence to
support the per-population statistics reported in
Sect.~\ref{sec:results}. Satellite-to-detection association inherits
the cross-matching pipeline of \citet{Grigg2025}, in which
detections are linked to NORAD catalogue entries via SGP4
propagation of publicly available two-line element sets. The
TLE positional uncertainty after a few hours of propagation is
typically of order arcminutes for Starlink-class LEO objects and
remains well below the EDA2 synthesised beam at these frequencies;
\citet{Grigg2025} report a cross-matching residual of order $1\%$
on the same input catalogue, which is the value adopted here. The
v2-Mini surface density in the EDA2 field of view is sufficiently
low that the residual misidentification rate is expected to remain
at the per-cent level, and cannot account for the population-level
differences of $\gtrsim30\%$ in median flux density and in
eclipse-state ratio reported in Sect.~\ref{sec:results}.
Table~\ref{tab:populations} summarises the three populations and
Fig.~\ref{fig:overview} shows their distribution-level properties.

\begin{table}[t]
  \caption{Starlink populations analysed in this work.}
  \label{tab:populations}
  \centering
  \begin{tabular}{lcrr}
    \hline\hline
    Population              & GCAT bus       & $N_{\rm sat}$ & $N_{\rm det}$ \\
    \hline
    v2-Mini DTC             & \texttt{V2MD}  &   $175$ &  $10\,180$ \\
    v2-Mini Ku (pre-Optical) & \texttt{V2M}  & $1\,623$ & $102\,197$ \\
    v1.x                    & \texttt{V1.0}/\texttt{V1.5} & $8$ &    $157$ \\
    \hline
    \multicolumn{2}{l}{Total (analysed)}     & $1\,806$ & $112\,534$ \\
    \hline
  \end{tabular}
  \tablefoot{$N_{\rm sat}$ is the number of distinct NORAD
    identifiers; $N_{\rm det}$ is the number of stacked detections
    at \texttt{fine\_channel\_index = 31} in the \citet{Grigg2025}
    catalogue. The v2-Mini Optical variant (\texttt{V2MO}) is
    absent from the catalogue.}
\end{table}

\subsection{Range-corrected flux density}
\label{sec:range}

To remove the geometric distance dependence that dominates the raw
flux densities, the range-corrected flux density is defined as
\begin{equation}
  S_{\rm norm} \equiv S_{\rm obs}\,(r_{\rm sat}/r_{\rm ref})^{2},
  \qquad r_{\rm ref} = 1000\,{\rm km},
  \label{eq:Snorm}
\end{equation}
where $S_{\rm obs}$ is the measured flux density in Jy and
$r_{\rm sat}$ is the instantaneous satellite range from EDA2 taken
from the TLE propagation in \citet{Grigg2025}. This normalisation
follows \citet{Bassa2024} and permits the direct comparison of
satellites observed at different geometries. Except where noted,
distribution-wise statistics are reported on the per-satellite
median of $S_{\rm norm}$, so that satellites with highly unequal
numbers of detections do not bias the comparison.

\subsection{Illumination geometry}
\label{sec:illum}

For each detection the sub-satellite latitude, longitude, and
altitude are derived by projecting the topocentric azimuth and
elevation reported by \citet{Grigg2025} to an Earth-centred
Earth-fixed (ECEF) frame at the EDA2 location, and then converting
to geodetic coordinates. To determine whether the satellite was in
sunlight or in the Earth's shadow at the time of the detection,
the apparent solar position in ECEF is computed from the
low-precision IAU solar model \citep{Seidelmann1992}, and a
cylindrical Earth-shadow approximation is then applied. A satellite
is considered illuminated if its position vector has a positive
component along the Sun direction, or if its perpendicular distance
to the Sun--Earth axis exceeds the mean Earth radius
$R_\oplus = 6378.137$\,km. This geometric test neglects atmospheric
refraction and umbra/penumbra effects, and is therefore appropriate
for the $300$--$412$\,km altitude range with non-trivial detection counts in which the v2-Mini
population orbits. The full set of coordinate transformations and
the cylindrical shadow test are given in Appendix~\ref{app:geom}.

\subsection{Statistical comparisons}
\label{sec:tests}

We perform three comparisons. First, the per-satellite median
range-corrected flux density of the DTC population is compared
with that of the Ku-only comparison population using the
Mann--Whitney U test, Cliff's $\delta$, and the ratio of medians
with $95\%$ bootstrap confidence intervals. The non-parametric
Mann--Whitney U test is adopted because the per-satellite flux
density distributions are strongly non-Gaussian and heavy-tailed
\citep[see][Chapter~5]{FeigelsonBabu2012} and exhibit the long
tails visible in Fig.~\ref{fig:hdtc-cdf}d. Cliff's $\delta$ is
reported alongside the $p$-value because the large sample sizes
render any small shift formally significant, so an explicit effect
size is required to judge practical relevance. Second, the fraction
of events detected in the XX polarisation feed is examined across
the $21$ coarse channels for deviations from an instrumental
baseline estimated from the pooled sample; a two-sided binomial
test is applied per channel, and the false discovery rate (FDR)
across the $21$ channels is controlled using the
Benjamini--Hochberg procedure \citep{Benjamini1995} at $q = 0.05$.
Third, the range-corrected flux density is compared between
illuminated and eclipsed passes, first on the pooled v2-Mini
sample and then separately on the DTC and Ku-only sub-samples.
The mathematical definitions of the test statistics are given in
Appendix~\ref{app:tests}.

\section{Results}
\label{sec:results}

\begin{figure*}[t]
  \centering
  \includegraphics[width=\textwidth]{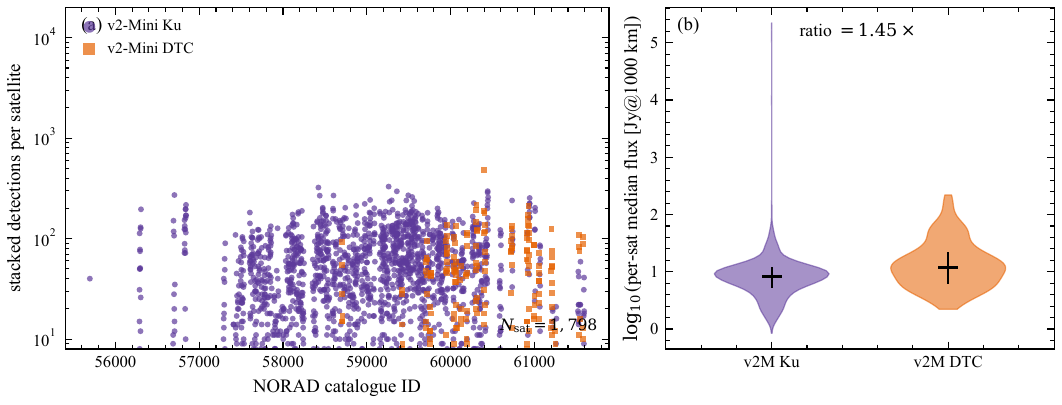}
  \caption{Overview of the $1\,798$ v2-Mini Starlink satellites in
    the \citet{Grigg2025} EDA2 catalogue after the GCAT
    classification and quality cuts of Sect.~\ref{sec:obs}.
    \textit{Left}: number of stacked detections per satellite as a
    function of NORAD catalogue identifier, for the two populations
    compared in this work.
    \textit{Right}: per-satellite median range-corrected flux
    density (in Jy at $1000$\,km), shown as a violin plot with
    interquartile range. The ratio of medians is quoted at the top
    of the panel.}
  \label{fig:overview}
\end{figure*}

\begin{figure*}[t]
  \centering
  \includegraphics[width=\textwidth]{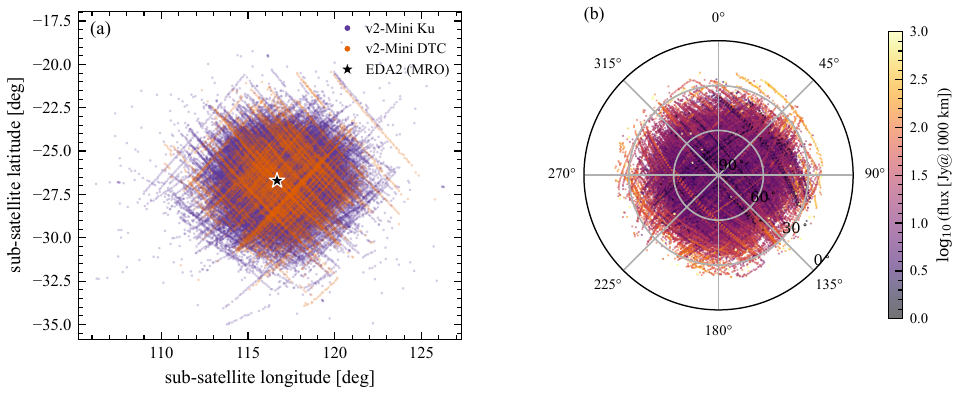}
  \caption{Pass geometry of the analysed v2-Mini detections.
    \textit{Left}: sub-satellite longitude and latitude for all
    v2-Mini Ku-only (purple) and DTC (orange) detections; the star
    marks the EDA2 location at the Murchison Radio-astronomy
    Observatory.
    \textit{Right}: azimuth and zenith-angle polar projection of
    all detections, colour-coded by $\log_{10}$ range-corrected
    flux density.}
  \label{fig:sky}
\end{figure*}

\begin{figure}[t]
  \centering
  \includegraphics[width=\columnwidth]{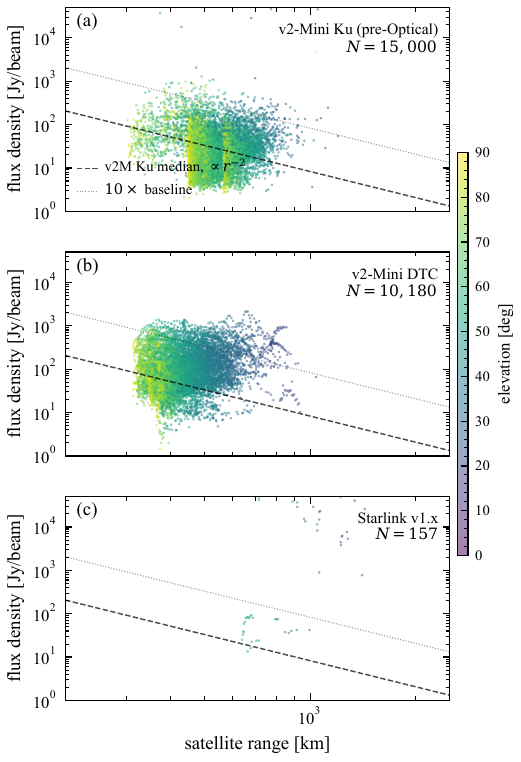}
  \caption{Observed flux density as a function of satellite range,
    colour-coded by elevation, for (a) v2-Mini Ku-only,
    (b) v2-Mini DTC, and (c) v1.x Starlink satellites. The dashed
    line in each panel shows the inverse-square scaling $\propto
    r^{-2}$ anchored at the median range-corrected flux density of
    the Ku-only comparison population, drawn as a guide to the eye;
    the dotted line shows ten times this baseline. Points offset
    above the dashed line correspond to UEMR in excess of the
    comparison median.}
  \label{fig:fluxrange}
\end{figure}

After applying the classification and quality cuts described in
Sect.~\ref{sec:obs}, the working sample contains $1\,798$ v2-Mini
Starlink satellites with a total of $112\,377$ range-corrected
detections. The v2-Mini DTC subsample accounts for $175$ satellites
and $10\,180$ detections, while the Ku-only pre-Optical v2-Mini
comparison accounts for $1\,623$ satellites and $102\,197$
detections. The eight v1.x satellites present in the catalogue
contribute only $157$ detections and are shown for reference in
Fig.~\ref{fig:fluxrange}c but are not used in the statistical
comparisons below. Figure~\ref{fig:overview}a shows that the
per-satellite detection count follows a broad distribution for both
populations; Fig.~\ref{fig:overview}b shows that their per-satellite
range-corrected flux density distributions overlap substantially,
with the DTC population shifted towards higher values.

Figure~\ref{fig:sky} shows that the two populations have essentially
overlapping sky coverage. Most detections occur in passes over the
Murchison Radio-astronomy Observatory at sub-satellite latitudes
between $-30^{\circ}$ and $-24^{\circ}$ and at satellite altitudes
between $350$ and $570$\,km. The azimuth/zenith-angle polar
projection (Fig.~\ref{fig:sky}b) is dominated by detections at
elevations above $45^{\circ}$, consistent with the sensitivity
pattern of the EDA2 station. The similarity of the two populations'
sky coverage is important for the statistical comparisons in
Sects.~\ref{sec:hdtc}--\ref{sec:hsmps}, because it ensures that any
differences in their flux density distributions cannot be attributed
to differences in pass geometry alone.

\begin{figure*}[t]
  \centering
  \includegraphics[width=\textwidth]{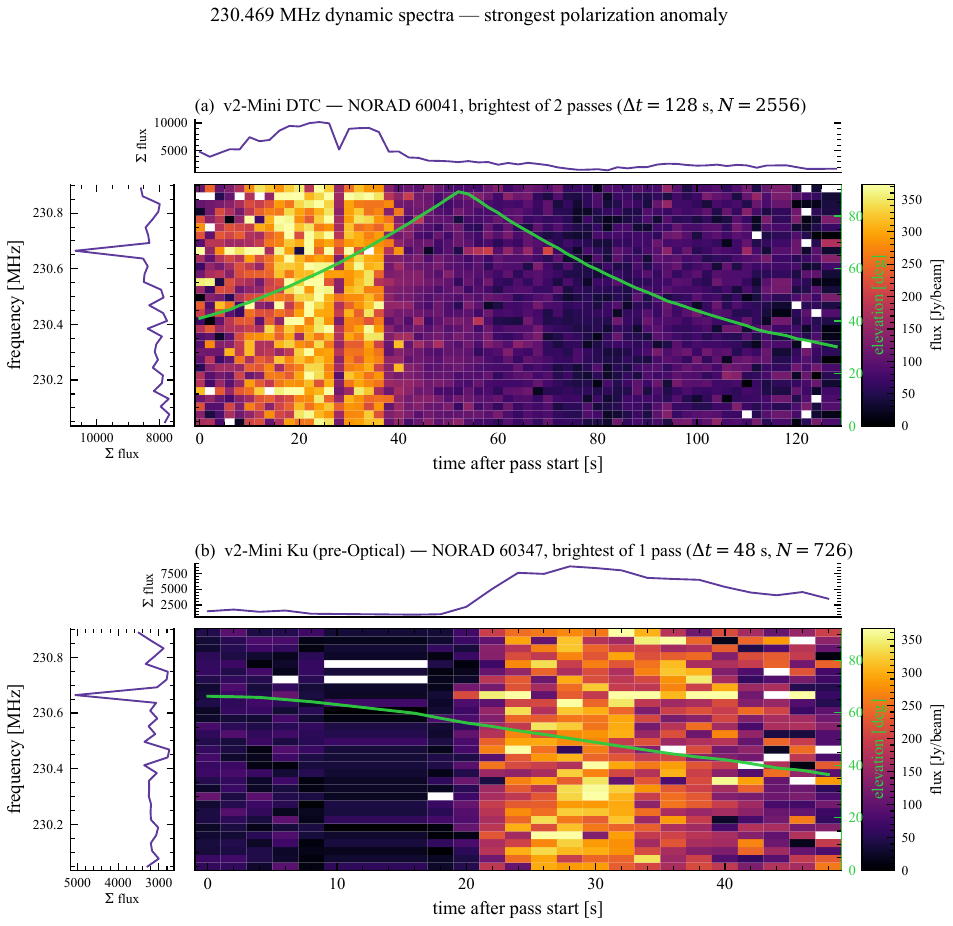}
  \caption{Dynamic spectra of the brightest pass at the
    $230.469$\,MHz coarse channel for (a) a representative v2-Mini
    DTC satellite (NORAD\,$60041$, pass duration $128$\,s) and
    (b) a representative Ku-only v2-Mini satellite
    (NORAD\,$60347$, pass duration $48$\,s), constructed from the
    $31$ fine channels within the coarse band. In each panel the
    brightest pass is the pass with the highest integrated
    $S_{\rm norm}$ at this coarse channel for that satellite. Top
    marginal panels show the flux density integrated across fine
    channels; left marginal panels show the time-integrated
    spectrum. The green curve overlays the satellite elevation
    during the pass. The two populations show qualitatively
    different time/frequency structure at this coarse channel,
    which is analysed statistically in Figs.~\ref{fig:hcmc} and
    \ref{fig:230deep}.}
  \label{fig:dynspec}
\end{figure*}

Figure~\ref{fig:dynspec} shows the dynamic spectrum of the brightest
pass at the $230.469$\,MHz coarse channel for one representative
DTC satellite (NORAD\,$60041$) and one representative Ku-only
comparison satellite (NORAD\,$60347$), constructed from the $31$
fine channels within the coarse band. The time--frequency structure
differs qualitatively between the two passes: the DTC pass shows a
concentrated flux density burst during the first $40$\,s, coincident
with the elevation rise from $40^{\circ}$ to $85^{\circ}$, while
the Ku-only pass shows its maximum flux density in the second half
of the pass as the elevation descends from $65^{\circ}$. Both show
emission that is broadband across all $31$ fine channels of the
coarse band. These dynamic spectra serve as motivation for the
statistical analysis of the $230.469$\,MHz channel in
Sect.~\ref{sec:hcmc}; an individual pair of passes cannot, on its
own, support a mechanism-level conclusion.

For visual clarity the Ku-only panel is randomly downsampled to
$15\,000$ detections (random seed $42$); all reported statistical
tests use the full $102\,197$-detection sample.

Figure~\ref{fig:fluxrange} shows the observed flux density as a
function of satellite range for the three populations, colour-coded
by instantaneous elevation. The dashed line in each panel is a
reference inverse-square envelope $\propto r^{-2}$ anchored at the
median range-corrected flux density of the Ku-only comparison
population. Panel~(a), which shows the Ku-only comparison, is
aligned with the dashed line by construction. Panel~(b), which
shows the DTC population, lies systematically above the dashed line
at all ranges, consistent with the per-satellite excess quantified
in Sect.~\ref{sec:hdtc}. Panel~(c), which shows the $157$ v1.x
detections, is sparse and is consistent with the baseline in the
range for which points exist, as expected if the range-corrected
statistics approximately align the v1.x and Ku-only v2-Mini
populations.

\subsection{DTC per-satellite excess}
\label{sec:hdtc}

\begin{figure*}[t]
  \centering
  \includegraphics[width=\textwidth]{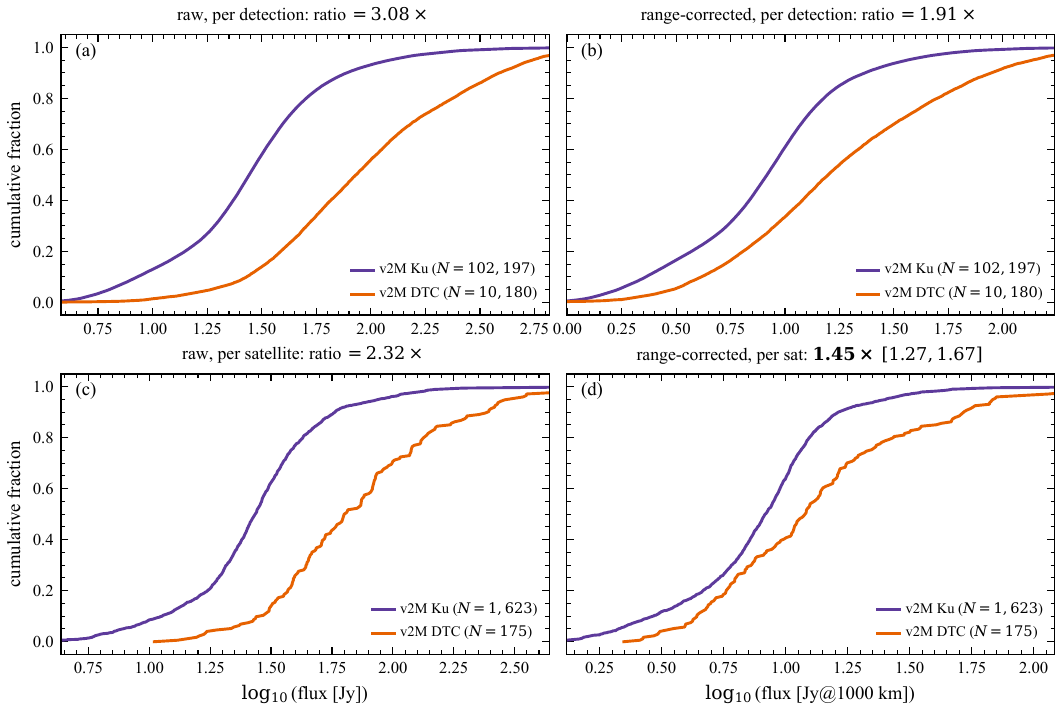}
  \caption{Cumulative distribution functions of the flux density for
    the DTC (orange) and Ku-only (purple) populations under four
    equivalent reductions:
    (a) raw per-detection;
    (b) range-corrected per-detection;
    (c) raw per-satellite median;
    (d) range-corrected per-satellite median.
    The ratio of medians is quoted in each panel title; the $95\%$
    bootstrap confidence interval is shown for panel~(d).}
  \label{fig:hdtc-cdf}
\end{figure*}

\begin{figure}[t]
  \centering
  \includegraphics[width=\columnwidth]{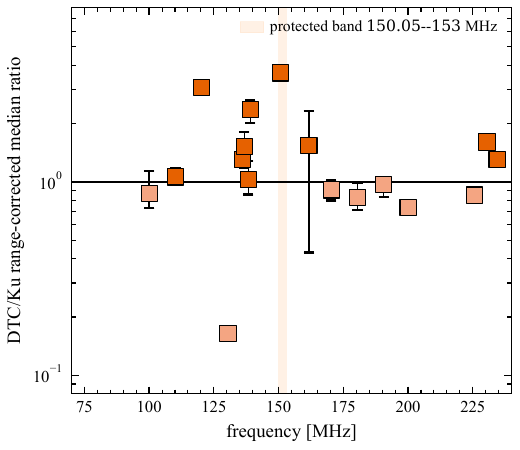}
  \caption{Per-frequency ratio of DTC to Ku-only range-corrected
    median flux density, with $95\%$ bootstrap confidence intervals.
    Squares filled in dark orange lie above unity (DTC-brighter
    channels); squares filled in lighter salmon lie below unity. The
    most extreme below-unity channel is at $130.469$\,MHz, where the
    DTC/Ku ratio falls to $\approx 0.16$. The shaded band indicates
    the Radio Astronomy Service band protected by the ITU between
    $150.05$ and $153$\,MHz.}
  \label{fig:hdtc-perfreq}
\end{figure}

Figure~\ref{fig:hdtc-cdf} shows the cumulative distribution function
of the flux density for the DTC and Ku-only comparison populations
under four equivalent reductions. The raw per-detection comparison
(panel a) is dominated by the range distribution of each population
and is therefore not a measurement of intrinsic brightness. The
range-corrected per-detection comparison (panel b) removes the
geometric factor and yields a DTC/Ku median ratio that is quoted in
the panel title. The per-satellite statistics in panels (c) and (d)
give equal weight to each satellite, and avoid biasing the
comparison in favour of satellites with unusually high detection
counts. The range-corrected per-satellite median (panel d) is
adopted as the main summary statistic for the per-satellite excess.

The main summary statistic yields a DTC/Ku median ratio of $1.449$
with a $95\%$ bootstrap confidence interval of $[1.27, 1.67]$ drawn
from $2000$ resamplings. A two-sided Mann--Whitney U test rejects
the null hypothesis of identical distributions at
$p = 2.6\times10^{-11}$. The effect size quantified by Cliff's
$\delta$ is $+0.303$. The direction of the effect is consistent
across the four reductions shown in Fig.~\ref{fig:hdtc-cdf}a--d,
and the ratio is largest for the per-satellite range-corrected
reduction, which is the reduction least affected by the geometry of
the observations. Cross-checking against the LOFAR sample of
\citet{Bassa2024}, the six DTC satellites observed there all appear
in our $175$-satellite DTC sample with the EDA2 detection counts
listed in Table~\ref{tab:bassa_xcheck}, confirming the cross-instrument
consistency of the per-satellite excess.

\begin{table}[t]
  \caption{Cross-instrument check against the LOFAR sample of
    \citet{Bassa2024}. All six DTC satellites observed there
    appear in our EDA2 sample with detection counts ranging from
    $16$ to $103$.}
  \label{tab:bassa_xcheck}
  \centering
  \footnotesize
  \begin{tabular}{rlr}
    \hline\hline
    NORAD ID & Starlink name & EDA2 detections \\
    \hline
    60039 & STARLINK-11140 & 54  \\
    60040 & STARLINK-11122 & 66  \\
    60041 & STARLINK-11149 & 103 \\
    60042 & STARLINK-11120 & 16  \\
    60043 & STARLINK-11086 & 18  \\
    60048 & STARLINK-11135 & 37  \\
    \hline
  \end{tabular}
  \tablefoot{Starlink names from the McDowell GCAT
    \citep{McDowell2020}; detection counts from this work's
    reanalysis of the \citet{Grigg2025} catalogue.}
\end{table}

The two datasets sample the same UEMR phenomenon at different
points in the time--frequency plane. The LOFAR observations of
\citet{Bassa2024} consist of two $1$-hr coherent COBALT
tied-array-beam acquisitions on 2024 July 19, with a native
resolution of $12.2$\,kHz $\times$ $41.94$\,ms across the LBA
($10$--$88$\,MHz) and HBA ($110$--$188$\,MHz) bands. The present
EDA2 reanalysis instead draws on the multi-night \citet{Grigg2025}
catalogue across $21$ discrete coarse channels spanning
$72.685$--$234.375$\,MHz, so the two observations overlap only in
the $110$--$188$\,MHz window.

Within that window, all six v2-Mini DTC satellites in the HBA sample
of \citet{Bassa2024} (NORAD $60039$--$60048$) appear in our
$175$-satellite DTC selection, with per-satellite EDA2 detection
counts ranging from $16$ to $103$. The dispersion arises from a
combination of pass geometry, the number of EDA2 transits per
satellite, and intrinsic UEMR variability, whose relative
contributions are not separable in the present sample. The range
is consistent in order of magnitude with the LOFAR flux densities
reported for the same six satellites in the $116$--$124$\,MHz band
($15.7$--$81.4$\,Jy at a satellite distance of $\sim 347$\,km).

The two methodologies are complementary. The $12.2$\,kHz LOFAR
resolution is fine enough to resolve the comb spacings of $50$ and
$150$\,kHz reported by \citet{Bassa2024} in the $116$--$124$\,MHz
band for v2-Mini DTC satellites; the EDA2 coarse channelisation
cannot. Conversely, the LOFAR HBA upper edge of $188$\,MHz lies
below the $230.469$\,MHz channel at which we report the largest
polarisation anomaly (Sect.~\ref{sec:hcmc}), and the spectral
structure at that frequency awaits confirmation from independent
broadband observations. The cross-check in
Table~\ref{tab:bassa_xcheck} is therefore best read as a
sample-membership consistency test; the per-satellite excess and
the \citet{Bassa2024} generation-over-generation factor of $32$
are related in Sect.~\ref{sec:discussion}.

Figure~\ref{fig:hdtc-perfreq} decomposes this comparison into each
of the $21$ coarse frequency channels. The majority of channels
show a DTC/Ku ratio above unity, consistent with the per-satellite
excess being broadband rather than concentrated in any single
frequency band. A single channel at $130.469$\,MHz shows the
opposite sense with a particularly extreme separation from unity,
where the DTC population is fainter than the Ku-only comparison by
a factor of approximately six (median ratio $0.16$). Several other
channels (e.g.\ $170$, $180$, $200$, $225$, $234$\,MHz; see
Table~\ref{tab:hdtc_perfreq_table}) show DTC median ratios modestly
below unity, but the magnitude of the $130.469$\,MHz separation
exceeds that of every other below-unity channel by an order of
magnitude. No physical interpretation of the $130.469$\,MHz
exception is attempted here.
The shaded band in Fig.~\ref{fig:hdtc-perfreq} marks the protected
Radio Astronomy Service band between $150.05$ and $153$\,MHz; the
one coarse channel that falls within this band is consistent with
the broadband trend and is not anomalous.

\begin{table}[t]
  \caption{Per-coarse-channel DTC/Ku range-corrected median flux
    ratio, complementing Fig.~\ref{fig:hdtc-perfreq}. Channels with
    fewer than $20$ DTC detections (four channels with
    $N_{\rm DTC}\in\{0,0,0,16\}$) are omitted because the per-channel
    median ratio is not reliably estimated; the $10\,164$ DTC
    detections tabulated here account for $10\,180-16=10\,164$ of
    the $10\,180$ DTC detections in the sample.}
  \label{tab:hdtc_perfreq_table}
  \centering
  \footnotesize
  \begin{tabular}{rrrrrr}
    \hline\hline
    $f$ [MHz] & $N_{\rm DTC}$ & $N_{\rm Ku}$ & DTC/Ku & KS $p$ & Cliff $\delta$ \\
    \hline
    $100.00$  &  $210$ &     $113$ & $0.87$ & $1.4\times10^{-2}$  & $+0.02$ \\
    $110.16$  &  $591$ &     $567$ & $1.06$ & $4.4\times10^{-3}$  & $+0.02$ \\
    $120.31$  & $1\,165$ &   $379$ & $3.08$ & $3.3\times10^{-90}$ & $+0.60$ \\
    $130.47$  &   $56$ &  $1\,638$ & $\mathbf{0.17}$ & $1.5\times10^{-69}$ & $-0.97$ \\
    $135.94$  &  $833$ &  $1\,458$ & $1.30$ & $1.1\times10^{-17}$ & $+0.14$ \\
    $136.72$  &  $944$ &  $2\,601$ & $1.52$ & $1.2\times10^{-38}$ & $+0.16$ \\
    $138.28$  &   $88$ &     $524$ & $1.03$ & $2.6\times10^{-1}$  & $+0.04$ \\
    $139.06$  &  $404$ &  $1\,443$ & $2.37$ & $1.5\times10^{-41}$ & $+0.42$ \\
    $150.78$  & $1\,113$ & $15\,766$ & $3.67$ & $<10^{-300}$ & $+0.77$ \\
    $161.72$  &   $71$ & $21\,984$ & $1.55$ & $5.4\times10^{-9}$  & $+0.06$ \\
    $170.31$  &  $261$ & $23\,242$ & $0.91$ & $2.2\times10^{-10}$ & $-0.11$ \\
    $180.47$  &  $140$ &  $5\,127$ & $0.83$ & $1.0\times10^{-5}$  & $-0.24$ \\
    $190.62$  &  $304$ &  $2\,587$ & $0.97$ & $1.2\times10^{-8}$  & $-0.14$ \\
    $200.00$  &  $623$ &  $6\,814$ & $0.74$ & $2.6\times10^{-27}$ & $-0.26$ \\
    $225.78$  &  $241$ &  $1\,643$ & $0.86$ & $1.9\times10^{-10}$ & $-0.25$ \\
    $230.47$  & $1\,765$ &   $939$ & $1.61$ & $3.4\times10^{-68}$ & $+0.40$ \\
    $234.38$  & $1\,355$ & $1\,270$ & $1.30$ & $2.2\times10^{-22}$ & $+0.22$ \\
    \hline
  \end{tabular}
  \tablefoot{$N_{\rm DTC}$ and $N_{\rm Ku}$ are the detection counts
    contributing to the per-channel median ratio. Cliff's $\delta$ is
    signed such that positive values indicate the DTC population is
    brighter; the only channel at which DTC is fainter at
    $|\delta|>0.5$ is $130.469$\,MHz (bold).}
\end{table}

\subsection{Coarse-channel polarisation anomalies}
\label{sec:hcmc}

\begin{figure}[t]
  \centering
  \includegraphics[width=\columnwidth]{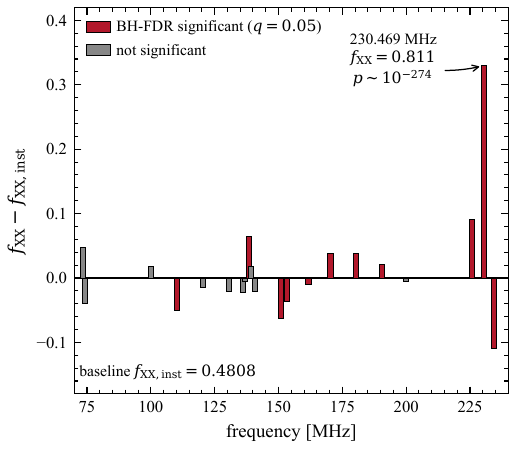}
  \caption{Deviation of the XX polarisation fraction from the
    instrumental baseline $f_{\rm XX,inst}$ in each coarse channel.
    Bars coloured in dark red survive the Benjamini--Hochberg false
    discovery rate control at $q = 0.05$; grey bars do not. The
    annotation at $230.469$\,MHz marks the channel with the largest
    deviation.}
  \label{fig:hcmc}
\end{figure}

\begin{figure*}[t]
  \centering
  \includegraphics[width=\textwidth]{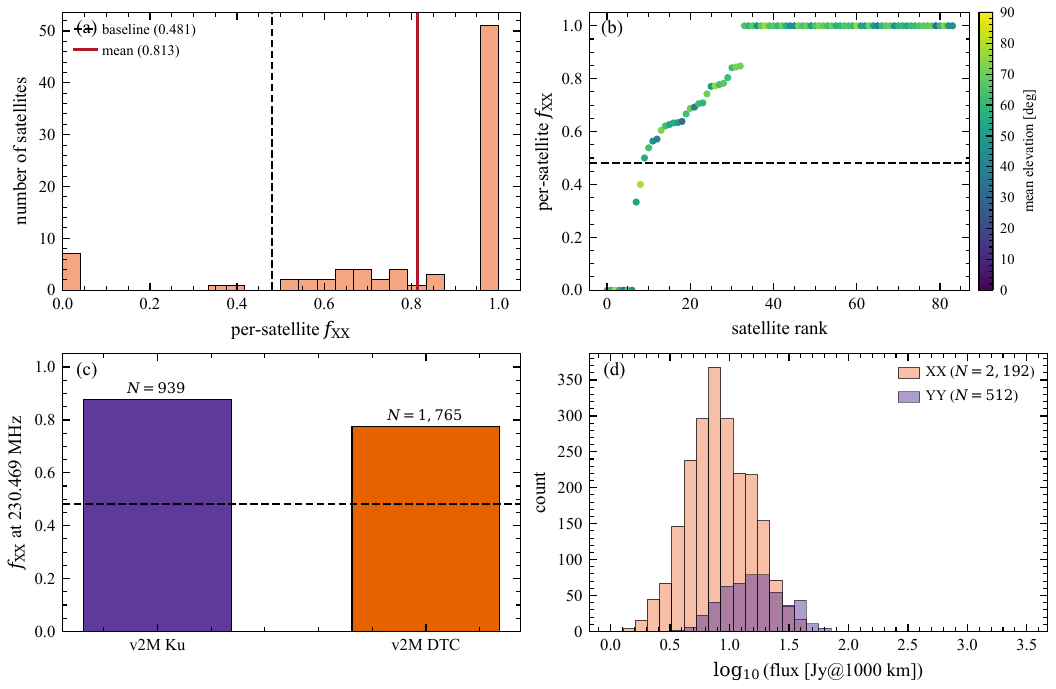}
  \caption{Characterisation of the $230.469$\,MHz XX fraction
    anomaly.
    (a)~Per-satellite XX fraction distribution for satellites with
    at least five detections in this channel.
    (b)~The same, sorted by XX fraction and colour-coded by mean
    elevation.
    (c)~XX fraction by population, with the number of detections
    quoted.
    (d)~Range-corrected flux density distributions in XX and YY,
    showing that the XX excess is not accompanied by a commensurate
    increase in the per-detection amplitude.}
  \label{fig:230deep}
\end{figure*}

Figure~\ref{fig:hcmc} shows the deviation of the XX polarisation
fraction from the instrumental baseline
$f_{\rm XX,inst} = 0.4809$ in each of the $21$ coarse channels. The
baseline is the pooled-sample XX fraction across the $112\,377$
v2-Mini detections analysed here, and represents the XX fraction
expected under the null hypothesis that the EDA2 orthogonal
polarisation feeds have equal sensitivity and the sources emit
unpolarised radiation; it is a pooled-data estimate rather than an
independent instrumental calibration, and the leave-one-channel-out
robustness check reported below confirms that the largest deviation
is preserved under a null baseline that explicitly excludes the
test channel. Positive bars indicate channels
in which the XX polarisation is observed more frequently than the
baseline, and negative bars indicate the opposite. A two-sided
binomial test is applied in each channel independently, and the
Benjamini--Hochberg FDR procedure at $q = 0.05$ is used to control
the false discovery rate across the $21$ tests. After this
correction, $11$ of the $21$ channels retain significance (coloured
bars in Fig.~\ref{fig:hcmc}), and the remaining $10$ are consistent
with the null baseline at the $q = 0.05$ level.

The largest deviation, by a wide margin, occurs at $230.469$\,MHz,
where the XX fraction reaches $f_{\rm XX} = 0.811$ with a $95\%$
Wilson score interval of $[0.795, 0.825]$. A two-sided binomial
test at this channel returns $p \sim 10^{-274}$, well beyond the
significance of any other channel in the dataset. The channel lies
$80$\,MHz above the protected Radio Astronomy Service band at
$150.05$--$153$\,MHz, and has no assigned Starlink service. The
binomial test used here addresses only the population-level
event-counting statistic; it does not address whether the emission
itself is linearly polarised in a physical sense, but only whether
EDA2 is more likely to detect the source in the X polarisation
feed than in the Y feed.

Figure~\ref{fig:230deep} characterises this channel in more detail.
Panel~(a) shows the distribution of per-satellite XX fractions for
satellites with at least five detections at $230.469$\,MHz; the
distribution is asymmetric and centred well above the baseline,
with most satellites showing $f_{\rm XX}$ between $0.7$ and $1.0$.
Panel~(b) sorts the same satellites by their XX fraction and
colour-codes each point by the mean observing elevation; no
elevation dependence is apparent, which argues against a dominant
beam-shape systematic. Panel~(c) shows that both the DTC and
Ku-only comparison populations contribute to the excess, with the
DTC population exhibiting the larger deviation, consistent with
the per-satellite excess reported in Sect.~\ref{sec:hdtc}.
Panel~(d) compares the log-flux distributions in the two
polarisations directly: the XX distribution contains substantially
more events than the YY distribution, but the two distributions
have similar shape and median, suggesting that the anomaly is in
the \emph{number} of detections rather than in the per-detection
amplitude.

Table~\ref{tab:hcmc_perfreq} summarises the per-channel binomial
test against the instrumental baseline and the Benjamini--Hochberg
significance status for all $21$ coarse channels in the dataset.

\begin{table}[t]
  \caption{Per-channel binomial test of the XX detection fraction
    against the pooled v2-Mini baseline $f_{\rm XX,inst}=0.4809$.
    Eleven of the $21$ coarse channels survive Benjamini--Hochberg
    false discovery rate control at $q=0.05$. The $230.469$\,MHz
    channel is the largest deviation in the dataset.}
  \label{tab:hcmc_perfreq}
  \centering
  \footnotesize
  \begin{tabular}{rrrrcc}
    \hline\hline
    $f$ [MHz] & $N$ & $f_{\rm XX}$ & $\Delta$ & $p$-value & BH \\
    \hline
     73.44 &   199 & 0.528 & $+0.047$ & $2\times10^{-1}$ &            \\
     74.22 &   344 & 0.442 & $-0.039$ & $2\times10^{-1}$ &            \\
    100.00 &   323 & 0.498 & $+0.018$ & $5\times10^{-1}$ &            \\
    110.16 &  1158 & 0.430 & $-0.051$ & $6\times10^{-4}$ & \checkmark \\
    120.31 &  1544 & 0.466 & $-0.015$ & $2\times10^{-1}$ &            \\
    130.47 &  1694 & 0.460 & $-0.020$ & $9\times10^{-2}$ &            \\
    135.94 &  2291 & 0.458 & $-0.023$ & $3\times10^{-2}$ &            \\
    136.72 &  3545 & 0.475 & $-0.006$ & $5\times10^{-1}$ &            \\
    138.28 &   612 & 0.554 & $+0.073$ & $3\times10^{-4}$ & \checkmark \\
    139.06 &  1847 & 0.499 & $+0.018$ & $1\times10^{-1}$ &            \\
    140.62 &   940 & 0.461 & $-0.020$ & $2\times10^{-1}$ &            \\
    150.78 & 16879 & 0.418 & $-0.063$ & $6\times10^{-61}$  & \checkmark \\
    153.12 & 12635 & 0.444 & $-0.037$ & $1\times10^{-16}$  & \checkmark \\
    161.72 & 22055 & 0.471 & $-0.010$ & $2\times10^{-3}$ & \checkmark \\
    170.31 & 23503 & 0.518 & $+0.038$ & $9\times10^{-31}$  & \checkmark \\
    180.47 &  5267 & 0.519 & $+0.038$ & $3\times10^{-8}$ & \checkmark \\
    190.62 &  2891 & 0.503 & $+0.022$ & $2\times10^{-2}$ & \checkmark \\
    200.00 &  7437 & 0.476 & $-0.004$ & $4\times10^{-1}$ &            \\
    225.78 &  1884 & 0.572 & $+0.091$ & $3\times10^{-15}$  & \checkmark \\
    230.47 &  2704 & 0.811 & $+0.330$ & $4\times10^{-275}$ & \checkmark \\
    234.38 &  2625 & 0.371 & $-0.109$ & $2\times10^{-29}$  & \checkmark \\
    \hline
  \end{tabular}
  \tablefoot{$N$: total detections in the channel; $f_{\rm XX}$:
    XX-feed detection fraction; $\Delta$: deviation from the
    pooled baseline; BH column marks channels surviving false
    discovery rate control at $q=0.05$ ($11/21$).}
\end{table}

As a robustness check against the circularity of estimating the
pooled-sample baseline from data that include the test channel, a
leave-one-channel-out baseline is reported. For each channel $c$, a
null XX fraction is computed from the other twenty channels combined,
and the binomial test of channel $c$ is repeated against this
channel-specific null. Under the leave-one-channel-out baseline,
$12$ of the $21$ channels survive Benjamini--Hochberg control at
$q=0.05$ (versus $11$ under the pooled-sample baseline), and the
$230.469$\,MHz channel returns a deviation of $+0.338$ against a
leave-one-out baseline of $0.473$ with a binomial $p \sim 10^{-287}$.
The qualitative pattern of which channels deviate, and in particular
the headline status of the $230.469$\,MHz anomaly, is therefore
preserved when the test channel is explicitly excluded from the
null baseline calibration.

\subsection{Spectral structure of detections}
\label{sec:spectra}

Two complementary spectral statistics summarise the per-channel
distribution of v2-Mini UEMR in the dataset. The first is the
detection-count column $N$ in Table~\ref{tab:hcmc_perfreq}, which
quantifies where the catalogue threshold is exceeded most often.
Detections cluster strongly at $150.78$, $153.12$, $161.72$, and
$170.31$\,MHz, which together account for $\sim 75\,000$ of the
$\sim 112\,500$ stacked detections in the analysed sample (about
two-thirds of the total). A secondary peak lies near $200$\,MHz,
and the $73$--$110$\,MHz lower band is sparsely populated. The
second statistic is the per-channel DTC/Ku ratio plotted in
Fig.~\ref{fig:hdtc-perfreq}: most channels show a ratio above
unity, with the largest single-channel exception at
$130.469$\,MHz, where the ratio falls to $\approx 0.16$ and the
DTC population is fainter than the Ku-only comparison.

The dataset can be sub-divided into three frequency regions for
discussion. The low-band region between $73$ and $110$\,MHz contains
only three coarse channels in the analysed sample, each with
$N < 350$ detections, and none of the three survives BH-FDR
correction; this region therefore contributes weakly to the
catalogue-level statistics and is consistent with sparse,
near-threshold emission. The mid-band region between $110$ and
$190$\,MHz contains eleven coarse channels and most of the
catalogue detections; it overlaps the LOFAR HBA window of
\citet{DiVruno2023} and \citet{Bassa2024}, in which v2-Mini DTC
satellites were observed to emit broadband UEMR plus narrow comb
structure at the $50$ and $150$\,kHz scales. The high-band region
between $200$ and $234.4$\,MHz contains four coarse channels
absent from the LOFAR HBA coverage; three of these four channels
($225.781$, $230.469$, and $234.375$\,MHz) survive BH-FDR
correction and together contain the largest XX-fraction excursions
in the dataset. The $230.469$\,MHz channel in particular sets the
most extreme polarisation anomaly observed in any v2-Mini
sub-population reported to date.

The per-channel asymmetry distribution above $200$\,MHz is therefore
not a simple continuation of the lower-frequency pattern. At
$225.781$ and $230.469$\,MHz the XX-fraction deviations have the
same positive sign ($+0.091$ and $+0.330$ respectively) but differ
by a factor of $\sim 4$ in amplitude; at $234.375$\,MHz the
deviation reverses sign and is the largest negative deviation in
the sample ($\Delta=-0.109$). No published Starlink Direct-to-Cell
band allocation overlaps the EDA2 frequency range, and the
per-channel structure cannot be attributed to direct in-band
emission from the cellular payload. Plausible origins include
harmonic and intermodulation products of on-board oscillators and
DC--DC converters, conducted emission from power-conditioning
subsystems, and out-of-band spectral leakage from communications
hardware. These mechanisms are not distinguishable from the EDA2
catalogue alone, because the EDA2 coarse-channel width does not
resolve the comb spacings of tens to hundreds of kHz reported by
\citet{Bassa2024} at LOFAR resolution. The mechanism discussion is
deferred to Sect.~\ref{sec:discussion}.

\subsection{Sub-MHz fine-channel structure of the $230.469$\,MHz anomaly}
\label{sec:finechan230}

The $230.469$\,MHz coarse-channel anomaly reported in
Sect.~\ref{sec:hcmc} is integrated over the full $\sim 0.78$\,MHz
coarse channel and is therefore agnostic to any sub-MHz spectral
structure. Because \citet{Bassa2024} resolved $50$ and $150$\,kHz
comb spacings within the $116$--$124$\,MHz band at LOFAR's
$12.2$\,kHz resolution, and because the $230.469$\,MHz channel lies
outside their HBA coverage, the natural follow-up is to inspect the
$\sim 24$\,kHz fine-channel data within the EDA2 coarse band.

\begin{figure*}[t]
  \centering
  \includegraphics[width=\textwidth]{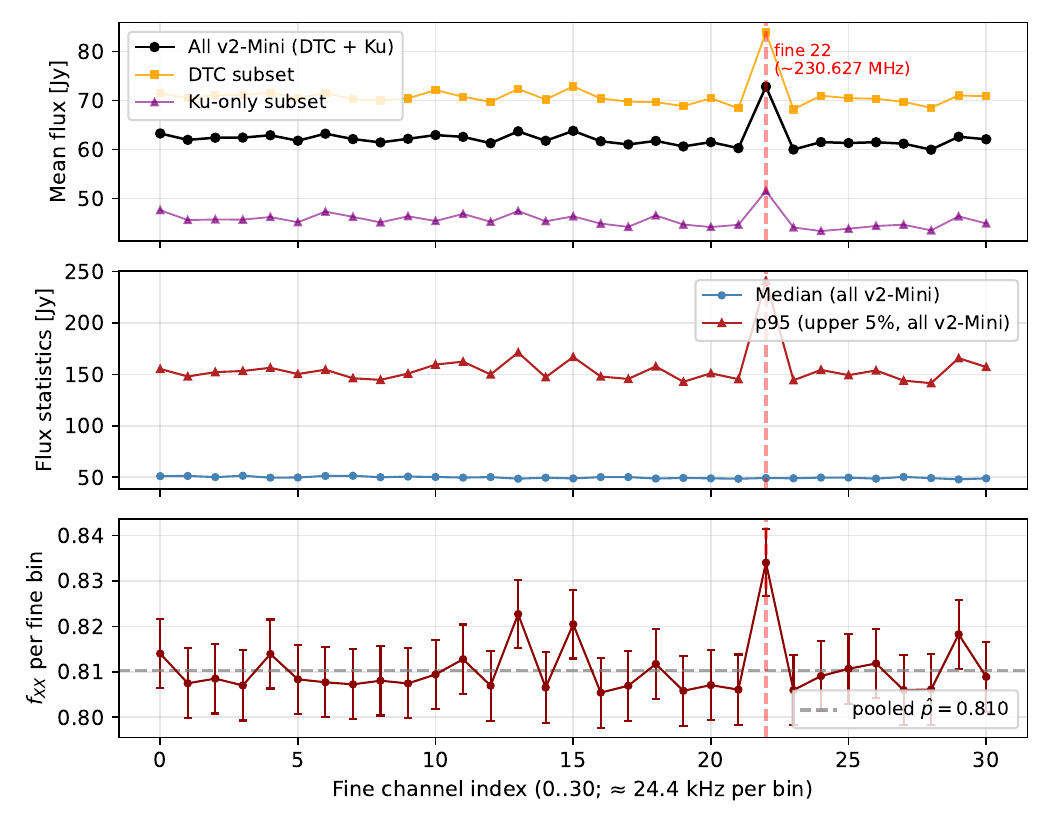}
  \caption{Per-fine-channel statistics within the $230.469$\,MHz
    coarse band ($N=80\,894$ v2-Mini fine-bin observations from the
    \citet{Grigg2025} catalogue, geometry-filtered). Top: mean flux,
    split DTC vs Ku-only. Middle: median and $95$th-percentile flux,
    pooled. Bottom: XX-feed fraction with binomial error bars; the
    pooled-sample value $\hat p=0.810$ is marked. Fine channel
    index $22$ ($\approx 230.627$\,MHz) is highlighted as the
    only bin with significantly elevated mean and upper-percentile
    flux; the XX-fraction variation across bins is constrained by
    the per-detection polarisation assignment in the catalogue and
    is informative only insofar as the mean shift reflects the
    larger-flux fraction at fine-bin $22$.}
  \label{fig:finechan230}
\end{figure*}

\begin{figure}[t]
  \centering
  \includegraphics[width=\columnwidth]{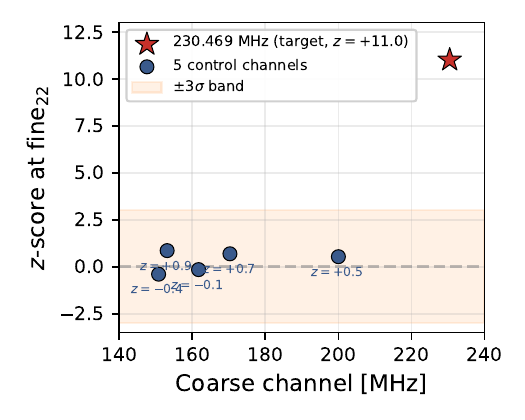}
  \caption{Cross-coarse-channel control for the fine-channel
    $22$ feature. The $z$-score is the deviation of the
    fine-channel-$22$ mean flux from the mean over the other
    $30$ fine bins, in units of the inter-bin standard deviation.
    The $230.469$\,MHz channel (red star) shows
    $z = +11.0$, whereas five high-occupancy control channels
    ($150.78$, $153.12$, $161.72$, $170.31$, and $200.00$\,MHz; blue
    circles) all lie within $\pm 1\sigma$, ruling out an
    instrumental fine-channel response artefact as the origin of
    the $230.627$\,MHz feature.}
  \label{fig:finechan_xchan}
\end{figure}

We isolate the $80\,894$ v2-Mini fine-bin observations
(fine-channel index $0$--$30$ in the \citet{Grigg2025} catalogue,
$\sim 24.4$\,kHz per bin) at the $230.469$\,MHz coarse channel and
recompute per-fine-bin statistics independently of the stacked
detection summary already used in Sect.~\ref{sec:hcmc}. As shown in
Fig.~\ref{fig:finechan230}, fine-channel index $22$
(corresponding to $\approx 230.627$\,MHz, an offset of $+159$\,kHz
from the coarse-channel centre) carries an empirical inter-bin
excess of $11.0$ standard deviations in the mean flux relative to
the other $30$ fine bins (top panel: $72.86$\,Jy at fine~$22$
versus $61.92 \pm 0.99$\,Jy elsewhere; per-bin sample sizes range
from $N=2\,525$ at fine~$22$ to $N=2\,641$ elsewhere, with the
$\sim 80\,894$ total v2-Mini fine rows reflecting the per-bin
post-filter rather than $2\,704 \times 31$ raw) and a corresponding
$+55\%$ excess in the $95$th-percentile flux (middle panel).
Applying a $31$-bin Bonferroni look-elsewhere correction places the
threshold at $z \approx 3.5$, well below the observed
$z = 11.0$. The median flux at fine~$22$ is not elevated, which
identifies the feature as a tail-driven phenomenon
contributed by a sub-population of bright detections rather than a
uniform shift in the bulk of the distribution.

The cross-coarse-channel control shown in
Fig.~\ref{fig:finechan_xchan} addresses a known concern:
EDA2 fine channels can in principle carry a fixed bandpass
artefact that would inflate flux at the same index across every
coarse channel. We re-run the per-fine-bin mean-flux estimator at
the five highest-occupancy control channels in the catalogue
($150.78$, $153.12$, $161.72$, $170.31$, and $200.00$\,MHz) and
recover $z$-scores between $-0.4$ and $+0.9$, all within the
$\pm 3\sigma$ band. The control therefore rules out a fixed
fine-channel-index bandpass artefact in the tested control
channels, and we attribute the $230.627$\,MHz excess to a
narrowband emission process specific to the $230.469$\,MHz coarse
channel rather than to a per-fine-index instrumental response;
adjacent high-band coarse channels ($225.781$ and $234.375$\,MHz)
are not used as controls because they themselves carry the largest
XX-fraction excursions in the catalogue
(Table~\ref{tab:hcmc_perfreq}) and cannot be assumed
feature-free.

Expressed as a ratio rather than absolute mean, the DTC subset
shows a fine-$22$ excess of $83.93 / 70.1 = 1.197\times$ over its
non-target bins, compared with the Ku-only ratio of
$\approx 50.2 / 46.8 = 1.073\times$, i.e.\ the DTC fine-channel
narrowband excess is approximately $0.12$ in fractional units
larger than the Ku-only excess at the same fine bin. The XX-feed
fraction at fine~$22$ ($0.834$) is nominally elevated above the
pooled per-bin value of $0.810$, but this comparison is not
independent: the \citet{Grigg2025} catalogue records a single
per-detection polarisation index that is propagated across all
$31$ fine-channel rows of the same detection, so per-fine-bin
XX/YY values are not separate measurements and the nominal
significance is not interpretable in the usual binomial sense.
The narrow flux excess therefore implies a narrowband emission
process at $\approx 230.627$\,MHz that contributes preferentially
to detections recorded in the XX feed; a definitive
mechanism-level interpretation is deferred to
Sect.~\ref{sec:mechanism}, where this fine-channel constraint is
combined with the eclipse-state result. Before that, the next
subsection (Sect.~\ref{sec:mechtests}) presents three falsifiable
yes/no tests that further constrain the physical origin of the
$230.627$\,MHz feature without invoking a phenomenological model.

\subsection{Falsifiable mechanism-discrimination tests}
\label{sec:mechtests}

\begin{figure*}[t]
  \centering
  \includegraphics[width=\textwidth]{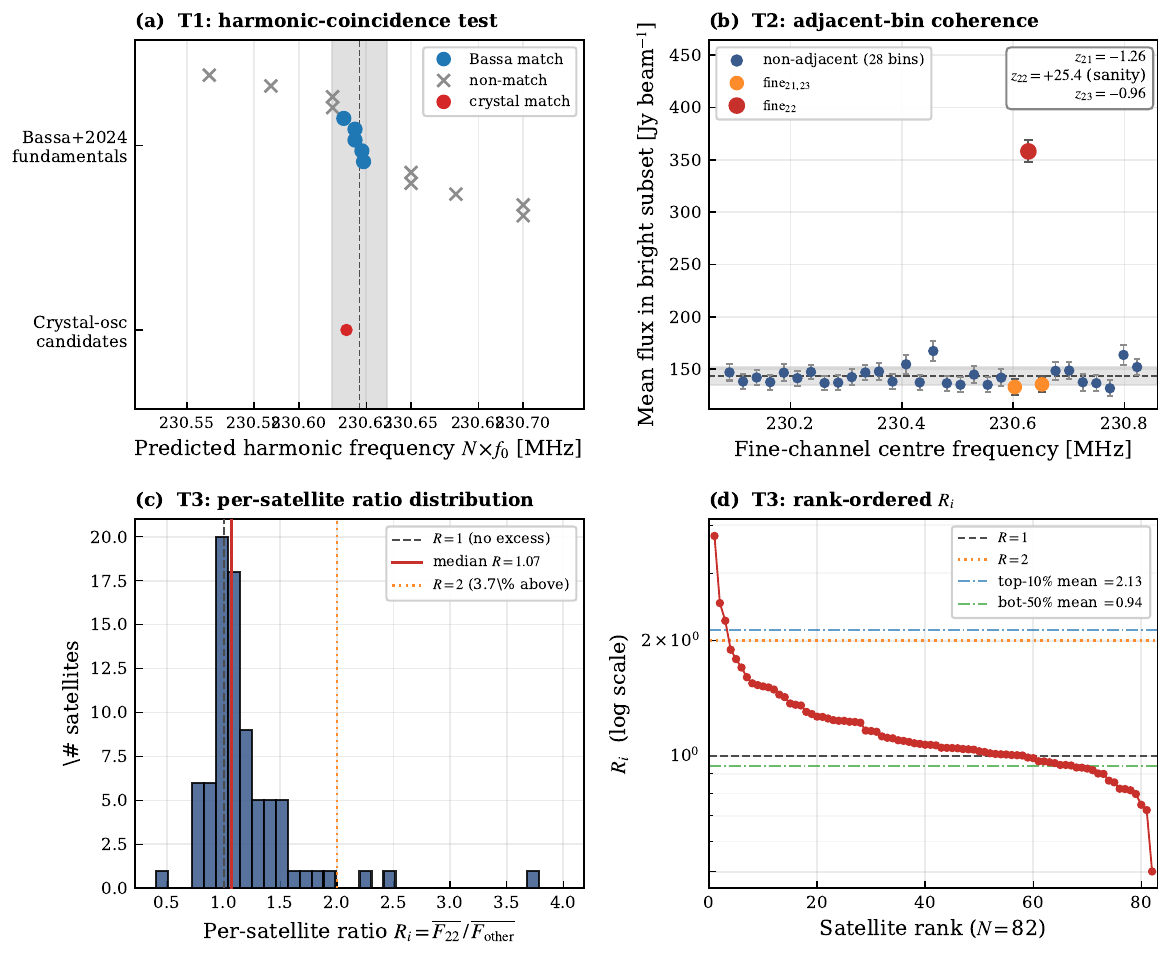}
  \caption{Three falsifiable mechanism-discrimination tests for the
    $230.627$\,MHz fine-channel feature reported in
    Sect.~\ref{sec:finechan230}. (a)~T1, harmonic-coincidence
    test: predicted frequencies $N \times f_0$ of all $14$
    fundamentals reported by \citet{Bassa2024} for v2-Mini Starlink
    satellites (top row) and a superset of $9$ plausible
    crystal-oscillator candidates (bottom row), evaluated at the
    integer $N$ minimising the residual relative to the target
    centroid $230.627441$\,MHz; the shaded band is the
    $\pm 12.207$\,kHz EDA2 half-fine-bin tolerance. Filled
    circles denote matches within tolerance and crosses denote
    non-matches. Five of the $14$ Bassa fundamentals produce a
    match within tolerance, compared with $5.73$ expected by chance
    under the null hypothesis $\sum_i \min(1, 2\Delta f/f_{0,i})$
    over the tested fundamentals. (b)~T2, adjacent-bin coherence
    within the bright subset (top $5\%$ by fine-$22$ flux,
    $N=127$ detections): the mean flux per fine bin is shown
    against the inter-bin baseline (shaded band,
    $\mu = 143.5$\,Jy\,beam$^{-1}$, $\sigma = 8.4$\,Jy\,beam$^{-1}$
    over the $28$ bins excluding fine-$21$, $22$, $23$). The
    fine-$22$ point at $z=+25.4$ is off-scale relative to
    the inter-bin spread; fine-$21$ ($z=-1.26$) and fine-$23$
    ($z=-0.96$) lie at or below baseline rather than above,
    falsifying the smearing hypothesis. (c)~T3, distribution of
    per-satellite ratios $R_i = \overline{F_{22}} /
    \overline{F_{\rm other}}$ across the $82$ v2-Mini satellites
    with $\ge 5$ detections at the $230.469$\,MHz coarse channel.
    (d)~Rank-ordered $R_i$ on a logarithmic axis; horizontal
    references mark the top-decile mean $R = 2.13$ and the
    bottom-half mean $R = 0.94$.}
  \label{fig:mech4panel}
\end{figure*}

The fine-channel narrowband excess of Sect.~\ref{sec:finechan230}
admits several physical interpretations -- harmonic emission from
on-board clock fundamentals, a switch-mode payload mode, an
intermodulation product between oscillators, uniform thermal
scaling across the population, or a hardware fault localised to a
small sub-population. Rather than fitting a multi-parameter
phenomenological model that can be tuned to match each of these,
we pose three falsifiable yes/no questions and design a numerical
test for each. The tests address: (T1) whether the $230.627$\,MHz
feature is consistent with an integer harmonic of any clock
fundamental resolved by \citet{Bassa2024} at LOFAR's
$12.2$\,kHz resolution; (T2) whether the fine-channel
excess is confined to a single $\sim 24.4$\,kHz bin or smears
into adjacent fine bins; and (T3) whether the per-satellite
distribution of fine-channel excess is dominated by a few
permanently-bright satellites, expressed uniformly across the
fleet, or distributed heterogeneously as a duty-cycle phenomenon.
The combined panel results are shown in
Fig.~\ref{fig:mech4panel}.

\paragraph{T1 -- Harmonic-coincidence test.}
We evaluated the integer-harmonic predictions $N \times f_0$ of
each of the $14$ Bassa+2024 clock fundamentals ($27.5$, $36.66$,
$55$, $110$, and $220$\,kHz for v2-Mini DTC LBA;
$37.5$, $50$, $75$ and $150$\,kHz for v2-Mini LBA;
$50$, $150$\,kHz for v2-Mini DTC HBA at $116$--$124$\,MHz;
$48.8$, $65$ and $97.5$\,kHz for v2-Mini HBA at $157$--$165$\,MHz)
and selected, for each fundamental, the integer $N$ that minimises
the residual relative to the EDA2 fine-channel-$22$ centroid
$230.627441$\,MHz. A match was declared when
$|N f_0 - 230.627441\,\mathrm{MHz}| \le 12.207$\,kHz, i.e.\ within
one half of the EDA2 fine bin (Fig.~\ref{fig:mech4panel}a). Five
of the $14$ fundamentals produce a match within tolerance. The
expected number of chance coincidences under the null hypothesis
of uniformly-distributed harmonics on the EDA2 fine grid is
$\sum_i \min(1, 2 \Delta f / f_{0,i}) = 5.73$ with
$\Delta f = 12.207$\,kHz, so the observed count is at the chance
level (observed/expected $= 0.87$) and the matching set is
scattered across fundamentals rather than concentrated on a
single one. The $50$ and $150$\,kHz fundamentals each appear
twice in the $14$-entry list (once for the LBA non-DTC v2-Mini
context and once for the HBA DTC v2-Mini context, in which
\citealt{Bassa2024} report the same numerical comb in distinct
band/population settings); treating these as a single oscillator
each reduces the expected count to $\approx 5.08$ and leaves the
observed/expected ratio at $0.98$ -- still at the chance level.
Because harmonically related fundamentals are not statistically
independent, the observed/expected ratio is reported as a
rule-out criterion rather than as a formal $p$-value. A superset of $9$ additional crystal-oscillator
candidates (10, 13, 16, 20, 25, 27, 100\,MHz and the
$12.288$\,MHz audio codec, plus a $32.768$\,kHz real-time clock)
yields one match (the $32.768$\,kHz RTC at harmonic $N = 7038$),
also at chance ($0.76$ expected). We conclude that the
$230.627$\,MHz feature is not uniquely tied to any
LOFAR-resolved Starlink clock fundamental, disfavouring a common
oscillator with the $116$--$165$\,MHz combs while not excluding
clock-spurious emission in general from other internal references
not characterised by the lower-band study.

\paragraph{T2 -- Adjacent-bin coherence.}
The single-bin nature of the fine-channel-$22$ excess identified
in Sect.~\ref{sec:finechan230} can be tested directly. Pivoting
the v2-Mini detections at the $230.469$\,MHz coarse channel into a
wide table indexed by fine-channel ($31$ fine bins per detection,
$N=2\,525$ detections with valid fine-$22$ flux after the
geometry filter), we restrict to detections in which the fine-$22$
flux exceeds its sample $95$th percentile ($p_{95} = 241$\,Jy,
$N = 127$ detections in the bright subset) and compute the mean
flux at each of the other $30$ fine bins within this subset
(Fig.~\ref{fig:mech4panel}b). The mean of the $28$ bins
excluding fine-$21$, fine-$22$, and fine-$23$ defines the
inter-bin baseline ($\mu = 143.5$\,Jy\,beam$^{-1}$,
$\sigma = 8.4$\,Jy\,beam$^{-1}$ across the $28$ bin means). The
$z$-scores relative to this baseline are $z_{22} = +25.4$ (a
sanity check confirming the bright cut), $z_{21} = -1.26$, and
$z_{23} = -0.96$; both adjacent bins therefore lie at or below
baseline rather than above. The smearing hypothesis -- in which
power leaks into adjacent fine bins from a feature broader than
$24.4$\,kHz -- requires an adjacent-bin elevation of
$z \gtrsim +3$ and is therefore falsified at the
$N = 127$ bright-subset sample size. The $230.627$\,MHz feature
is consistent with confinement to a single $24.4$\,kHz bin: the
intrinsic line width is unresolved at the EDA2 fine-channel
resolution and any sharper measurement requires a higher-resolution
follow-up.
The negative adjacent-bin $z$-scores are not individually
significant but lie in the direction expected for a narrowband
emitter that concentrates power into one bin and leaves the
neighbouring bins at the population mean or marginally below.

\paragraph{T3 -- Per-satellite ratio distribution.}
For each of the $82$ v2-Mini satellites with $\ge 5$ detections
at the $230.469$\,MHz coarse channel, we computed the per-satellite
ratio
$R_i = \overline{F_{22,i}}\, /\, \overline{F_{\mathrm{other},i}}$
in which the numerator is the satellite's mean fine-$22$ flux and
the denominator is its mean over the $28$ non-adjacent fine bins.
The distribution (Fig.~\ref{fig:mech4panel}c) has median
$R = 1.07$, mean $R = 1.18$, $p_{95} = 1.79$, and maximum
$R = 3.75$. We use this distribution to test two disjoint
mechanism alternatives. First, the \emph{hardware-fault}
hypothesis posits that the population-level excess is driven by a
small sub-population of permanently-bright units (e.g.\ payload
faults). Three of the $82$ satellites ($3.7\%$) have $R_i > 2$.
Re-running the same inter-bin $z$-score machinery applied in
Sect.~\ref{sec:finechan230} on the $79$ remaining satellites (i.e.\
removing all detections from the three $R_i > 2$ satellites and
recomputing the per-fine-bin means and inter-bin baseline) yields
$z_{22} = +12.95$ -- the excess survives at high significance once
the brightest three are removed, so a localised hardware fault
alone does not account for the signal. Second, the
\emph{uniform-population} hypothesis posits that every v2-Mini
satellite contributes the same fractional excess. The mean of the
top decile of $R_i$ is $2.13$, whereas the mean of the bottom half
is $0.94$ (Fig.~\ref{fig:mech4panel}d); the $2.26\times$ ratio
strongly disfavours a uniform shift, and the $45/82$ ($54.9\%$)
satellites with $R_i > 1.05$ show that the excess is neither
limited to a few units nor uniformly expressed. The two extremes
are thus strongly disfavoured, and the data favour a heterogeneous
duty-cycle expression across the v2-Mini fleet.

\paragraph{Combined verdict.}
The three tests jointly constrain the mechanism without
identifying it. The $230.627$\,MHz feature (T1) is not coincident
with any uniquely-identified LOFAR-resolved clock fundamental,
(T2) is confined to a fine-channel width $\le 24.4$\,kHz at the
EDA2 resolution, and (T3) is expressed heterogeneously across the
v2-Mini population rather than as a uniform thermal scaling or a
localised hardware fault. The combined constraints are consistent
with an intermittent narrowband emission whose amplitude depends
on the operational state of the spacecraft -- for example a
payload-active, communications-active, or power-management regime,
a switch-mode power-supply harmonic that activates with service
pattern, or an intermodulation product between on-board oscillators
not catalogued by \citet{Bassa2024} -- and these constraints apply
to the v2-Mini population as a whole rather than only to the DTC
subset, since the $230.627$\,MHz fine-channel feature is present in
both the DTC and Ku-only subpopulations (Sect.~\ref{sec:finechan230}). but the present data cannot uniquely
identify the emitting subsystem. A mechanism-level confirmation
would require telemetry-correlated observation cadence or direct
payload characterisation, neither of which is available in the
public domain at present. The orthogonal eclipse-state evidence
considered next (Sect.~\ref{sec:hsmps}) provides an independent
handle on the same question.

\subsection{Eclipse-state dependence}
\label{sec:hsmps}

\begin{figure*}[t]
  \centering
  \includegraphics[width=\textwidth]{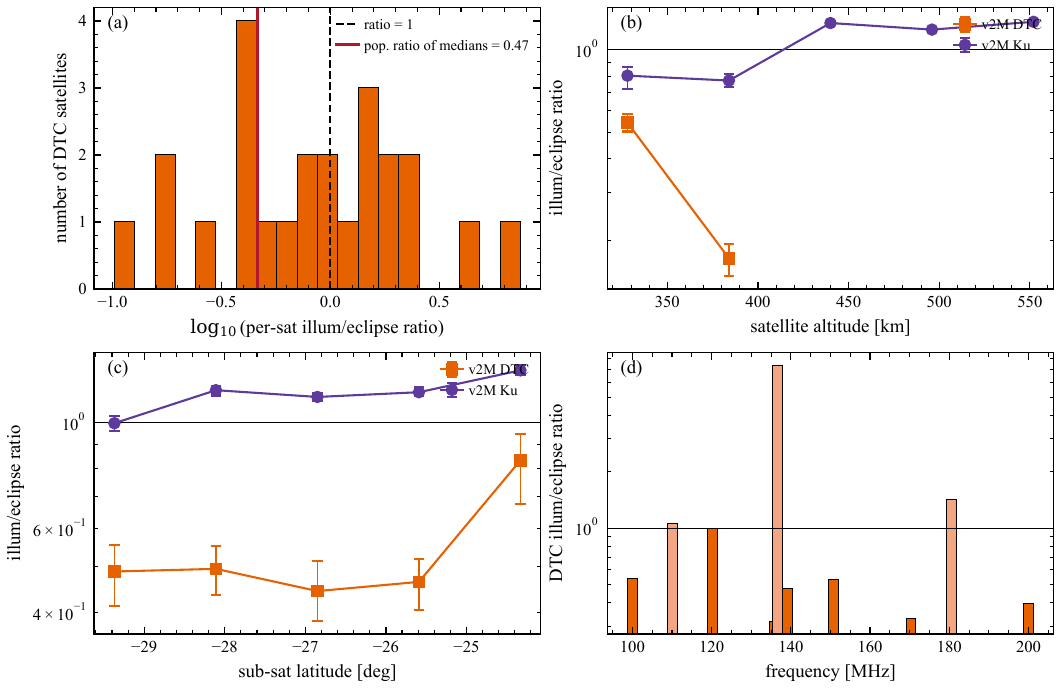}
  \caption{Robustness of the DTC eclipse-state reversal.
    (a)~Distribution of per-satellite illuminated/eclipsed ratios
    for DTC satellites with at least five detections in each state;
    the solid vertical line marks the population ratio of medians
    (the same statistic adopted in the primary test in the text),
    and the dashed line marks unity.
    (b)~Illuminated/eclipsed ratio as a function of satellite
    altitude, for DTC (orange squares) and Ku-only (purple circles)
    populations.
    (c)~Ratio as a function of sub-satellite latitude.
    (d)~Ratio as a function of frequency, for DTC only. Channels
    with fewer than $30$ illuminated or $30$ eclipsed detections
    are omitted; in particular, $230.469$\,MHz is absent because
    no eclipsed detections were recorded in this channel during
    the observing campaign.
    Error bars show $95\%$ bootstrap confidence intervals; the
    horizontal line at unity marks the null hypothesis.}
  \label{fig:eclipse}
\end{figure*}

The pooled-population form of the eclipse-state comparison is
tested first, in which the range-corrected flux density of
\emph{all} v2-Mini satellites is compared between illuminated and
eclipsed states without regard to the payload variant. The pooled test returns an illuminated/eclipsed ratio of medians of
$1.164\,[1.151,1.177]$, formally rejecting the null of no dependence but by a small
margin. Taken on
its own, this result would suggest a weak positive correlation
between illumination and UEMR, consistent with a solar-array-driven
power-conditioning mechanism. As shown below, however, this pooled
result is a pooling artefact of the unequal sub-population sizes
combined with opposite-signed within-population trends, and its
physical interpretation after sub-population separation is
misleading.

When the same test is applied separately to the DTC and Ku-only
comparison populations, the picture changes qualitatively. The
Ku-only comparison population yields an illuminated/eclipsed ratio
of $1.184\,[1.172, 1.197]$ ($95\%$ bootstrap confidence interval),
consistent in direction with the pooled
result and in the expected sense: illuminated Ku-only satellites
emit more UEMR than eclipsed ones. The DTC population, by contrast,
yields an illuminated/eclipsed ratio of $0.465\,[0.439, 0.492]$
($95\%$ bootstrap confidence interval), corresponding to DTC
satellites emitting roughly $2.1\times$ more
UEMR when eclipsed than when illuminated. The two population ratios have opposite signs, and their $95\%$
confidence intervals are widely disjoint, implying that they differ
at high significance under the cluster-robust restatement reported later in this section (joint satellite-level bootstrap, two-sided $p = 5\times 10^{-4}$). Because
the Ku-only comparison contributes an order of magnitude more
detections than the DTC population, the Ku-only sign dominates the
pooled median and masks the DTC reversal.

Figure~\ref{fig:eclipse} examines the robustness of the DTC
eclipse-state reversal against the most obvious systematics.
Panel~(a) shows the distribution of per-satellite
illuminated/eclipsed ratios for DTC satellites with at least five
detections in each state. The distribution extends predominantly
below unity; the solid vertical line in the panel marks the
population ratio of medians, $0.47$, which is the same summary
statistic adopted in the primary test above and is \emph{not} the
same quantity as the median of the per-satellite ratios, which is
closer to unity because individual satellites with modest
detection counts scatter around the population trend. Panel~(b)
shows the ratio as a function of satellite altitude in bins of
$56$\,km; the DTC reversal is preserved across the
$300$--$412$\,km altitude range with non-trivial detection counts, while the Ku-only comparison
ratio remains above unity in every altitude bin. The numerical
content of the populated DTC altitude bins is summarised in
Table~\ref{tab:eclipse_by_altitude}.

\begin{table}[t]
  \caption{Per-altitude-bin DTC illuminated/eclipsed range-corrected
    median flux ratio, complementing Fig.~\ref{fig:eclipse}b. Bins
    above $412$\,km contain fewer than five detections in each state
    and are omitted.}
  \label{tab:eclipse_by_altitude}
  \centering
  \footnotesize
  \begin{tabular}{lrrr}
    \hline\hline
    Altitude bin [km] & $N_{\rm illum}$ & $N_{\rm eclip}$ & illum/eclip ratio \\
    \hline
    $300$--$356$ & $4\,998$ & $2\,076$ & $0.542$ \\
    $356$--$412$ & $2\,545$ & $\phantom{0\,}117$ & $0.173$ \\
    \hline
  \end{tabular}
  \tablefoot{Both populated bins yield a ratio well below unity,
    consistent with the population-level DTC reversal reported in
    Sect.~\ref{sec:hsmps}. Bins above $412$\,km are sparsely populated
    in the present sample.}
\end{table}

Panel~(c) shows the same comparison as a function of sub-satellite
latitude. The analysed sample is concentrated in a narrow latitude
strip, but the DTC ratio remains below unity in every latitude bin,
which argues against a geographic selection effect coupled to the
day/night cycle through the local solar time. The bin centred at
$-24.3^\circ$ contains the smallest number of eclipsed DTC
detections ($N_{\rm ecl}=286$, against $414$--$583$ in the other
bins) and is correspondingly the noisiest, with a ratio of
approximately $0.85$; the four other bins lie between $0.45$ and
$0.55$.
Panel~(d) shows the DTC ratio as a function of the coarse
frequency channel; the reversal is preserved in the majority of
channels where the sample size permits a measurement, although a
small number of channels show ratios consistent with unity.

A complementary robustness check resamples at the satellite level
rather than at the detection level: bootstrapping over the $175$ DTC
satellites with replacement and re-evaluating the population ratio
of medians on each resample yields a $95\%$ confidence interval of
$[0.338, 0.627]$, wider than the detection-level interval
$[0.439, 0.492]$ as expected from the smaller effective sample size,
but still entirely below unity. Table~\ref{tab:eclipse_summary}
summarises the eclipse-state ratios across all populations and
sample variants considered in this section.

\begin{table*}[t]
  \caption{Illuminated/eclipsed range-corrected flux density ratios
    for the four populations and sample variants considered in
    Sect.~\ref{sec:hsmps}. Confidence intervals are detection-level
    bootstrap with $B=2000$ resamplings unless otherwise noted.}
  \label{tab:eclipse_summary}
  \centering
  \footnotesize
  \begin{tabular}{lrrl}
    \hline\hline
    Population & $N_{\rm sat,\,illum}/N_{\rm sat,\,ecl}$ & $N_{\rm det,\,illum}/N_{\rm det,\,ecl}$ & illum/ecl ratio \\
    \hline
    Pooled v2-Mini                & ---       & 81\,922 / 30\,455 & $1.164\,[1.151,1.177]$ \\
    Full DTC                      & 159 / 40  & 7\,686 / 2\,494   & $0.465\,[0.439,0.492]$ \\
    Full Ku-only                  & 1402 / 703 & 74\,236 / 27\,961 & $1.184\,[1.172,1.197]$ \\
    Matched Ku-only               & 896 / 422 & 50\,297 / 17\,999 & $1.188\,[1.169,1.209]$ \\
    DTC, satellite-bootstrap      & 159 / 40  & ---               & $0.465\,[0.338,0.627]$ \\
    Ku-only, satellite-bootstrap  & ---       & ---               & $1.184\,[1.115,1.257]$ \\
    DTC/Ku ratio of ratios        & ---       & ---               & $0.393\,[0.282,0.532]$ \\
    \hline
  \end{tabular}
  \tablefoot{Matched Ku-only restricts the comparison to the
    $998$ Ku-only satellites launched within the
    2024-01-03--2024-10-18 window spanned by the DTC sample;
    discarding the $625$ earlier-launched Ku-only satellites.
    The satellite-bootstrap row resamples DTC satellites with
    replacement rather than detections, yielding a wider but
    still entirely sub-unity interval.}
\end{table*} 

The stratified comparisons in
Fig.~\ref{fig:eclipse}b--d are reported as descriptive robustness
checks rather than independent hypothesis tests, and we do not apply
additional multiple-comparisons control to them; the headline test
remains the $95\%$ confidence intervals of the DTC and Ku-only
eclipse ratios under satellite-level cluster bootstrap, which are
widely disjoint and yield a two-sided interaction
$p = 5\times 10^{-4}$.

A potential confound is that the DTC and Ku-only populations were
deployed in different launch epochs, so the reversal could in
principle reflect satellite age or commissioning state rather than
the DTC payload. To test this, we restrict the Ku-only comparison
to the $998$ satellites whose launch dates fall within the
2024 January--October window spanned by the $175$ DTC satellites,
discarding the $625$ earlier-launched Ku-only satellites that have
no DTC counterpart. The eclipse-state ratio of the matched Ku-only
sub-sample is $1.188\,[1.169, 1.209]$, statistically indistinguishable
from the full Ku-only ratio of $1.184\,[1.172, 1.197]$, while the
DTC ratio is unchanged at $0.465\,[0.438, 0.494]$ by construction.
The two matched-sample confidence intervals remain disjoint with an
interaction $p$ of order $5\times 10^{-4}$ under the same
satellite-level cluster bootstrap. The reversal therefore is not an
artefact of the launch-epoch difference between the two populations.

The reversal also survives a shared-systematics argument. The DTC
and Ku-only populations are processed through identical EDA2
calibration, ephemeris, and flux-density pipelines; any geometric,
calibration, or pipeline-level systematic that produces the
Ku-only illuminated/eclipsed ratio of $1.184$ would drive the DTC
population in the same direction. The observed sign reversal at
$0.465$ is therefore incompatible with a shared-pipeline artefact,
independent of the matched-launch and altitude-bin checks reported
above.

The detection-level confidence intervals reported above treat
individual detections as approximately independent, which is
violated by the satellite/pass/frequency clustering of the EDA2
catalogue. A cluster-robust restatement resamples at the satellite
level ($N_{\rm sat,\,DTC} = 175$, $N_{\rm sat,\,Ku} = 1\,623$), pools
detections within each resample, and recomputes the
illuminated/eclipsed median ratio. The cluster bootstrap yields a
DTC ratio of $0.465\,[0.338, 0.627]$ and a Ku-only ratio of
$1.184\,[1.115, 1.257]$, both consistent in direction and central
value with the detection-level estimates but with intervals widened
to reflect the within-satellite correlation. A joint cluster
bootstrap of the difference, drawing the DTC and Ku-only satellite
samples independently, gives
$({\rm ratio}_{\rm DTC} - {\rm ratio}_{\rm Ku}) = -0.72\,[-0.87, -0.54]$
and a ratio of ratios of $0.39\,[0.28, 0.53]$, with a two-sided
bootstrap $p = 5\times 10^{-4}$ under the null of equal eclipse-state
ratios across payload classes. The cluster-robust restatement
therefore retains the reversal at high significance while widening
the formal interval by roughly a factor of three relative to the
detection-level $0.465\,[0.439, 0.492]$ estimate, and we adopt the
cluster-robust $p$ as the headline.

For the operational planning of long-duration SKA-Low observations,
the practical implication of the reversal is the time-averaged
contribution of DTC satellites integrated across the two
illumination states. Of the $10\,180$ DTC detections analysed
here, $7\,686$ ($75.5\%$) were recorded while the satellite was
illuminated and $2\,494$ ($24.5\%$) while eclipsed. Combined with
the population ratio of $1/0.465 \approx 2.15$ between eclipsed
and illuminated range-corrected flux densities, the time-averaged
DTC contribution is approximately $1.28$ times larger than the
illuminated-only estimate alone; a mitigation model calibrated
against illuminated DTC passes would correspondingly under-predict
the integrated UEMR at SKA-Low frequencies by roughly $28\%$, under
the working assumption that the per-state detection rate scales
linearly with the per-state exposure time; this assumption is
consistent with the EDA2 pipeline but is not directly verified, so
the under-prediction factor is presented as an illustrative
estimate rather than a calibrated correction.

The frequency-dependent decomposition in Fig.~\ref{fig:eclipse}d
shows that the DTC reversal is not concentrated in any single
channel. The reversal direction is preserved at the four
highest-$N$ channels in the sample ($150.78$, $153.12$, $161.72$,
and $170.31$\,MHz), all of which lie within the LOFAR HBA window
where \citet{Bassa2024} reported strong broadband v2-Mini UEMR.
The reversal is therefore a population-level property of the DTC
payload class across the dominant emission band of the catalogue,
rather than a feature of a single anomalous channel; in particular,
it is not driven by the $230.469$\,MHz polarisation excess
discussed in Sect.~\ref{sec:hcmc}.

A further internal consistency check examines the per-satellite
distribution rather than the population aggregate. Among the $24$
DTC satellites with at least five detections in each illumination
state, the median per-satellite illuminated/eclipsed ratio is
$0.886\,[0.441, 1.598]$ ($95\%$ cluster bootstrap, $N=24$); the
interquartile range $[0.42, 1.62]$ is asymmetric about unity, with
the lower quartile below unity by a larger margin than the upper
quartile is above. Fourteen of the $24$ satellites ($58\%$) lie
below unity and ten ($42\%$) below $0.75$. The per-satellite
distribution is therefore directionally consistent with the
detection-level reversal but the per-satellite cluster bootstrap
interval crosses unity, reflecting the modest sample size of DTC
satellites with reliable two-state coverage; the central value is
several times closer to unity than the detection-level
$0.465\,[0.439,0.492]$ estimate, indicating that the headline
ratio is driven primarily by a tail of heavily-detected
strongly-eclipsed-louder satellites rather than by uniform
suppression in the illuminated state.

\begin{figure*}[t]
  \centering
  \includegraphics[width=\textwidth]{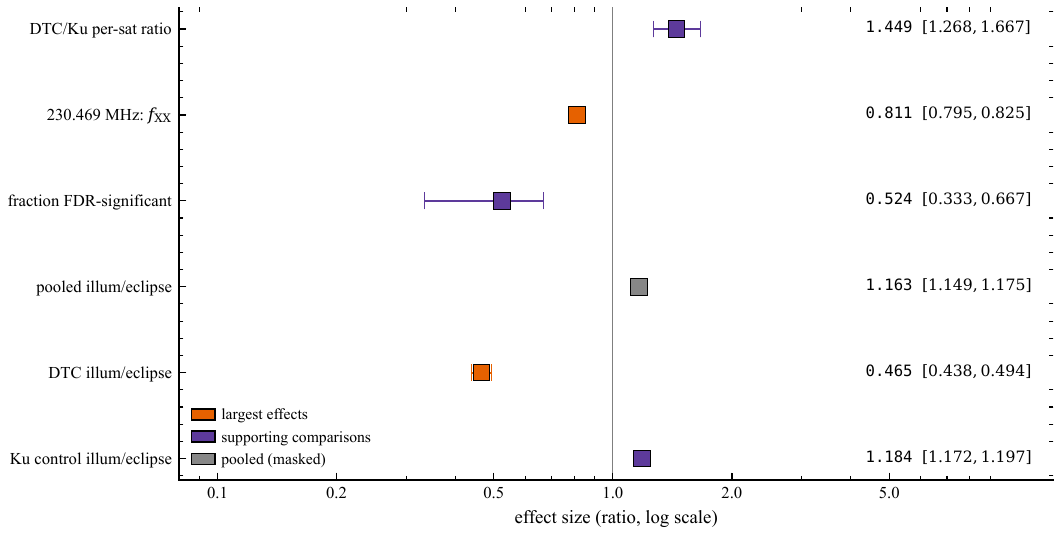}
 \caption{Summary of the effect sizes reported in
    Sect.~\ref{sec:results}. Intervals show $95\%$ bootstrap
    confidence intervals for median ratios and $95\%$ Wilson
    score intervals for binomial proportions; the plotted
    interval for the Benjamini--Hochberg FDR-significant channel
    fraction is a visual guide and not a formal interval. This
    panel combines ratios and proportions on a common numerical
    axis as a compact visual summary, not as commensurate effect
    measures. Orange symbols mark the two largest effects, namely
    the $230.469$\,MHz XX fraction anomaly and the DTC
    eclipse-state reversal. Purple symbols mark the supporting
    comparisons: the DTC/Ku per-satellite excess, the fraction of
    channels surviving Benjamini--Hochberg FDR control, and the
    Ku-only illuminated/eclipsed ratio. The grey symbol marks the
    pooled v2-Mini illuminated/eclipsed test, which is dominated
    by the larger Ku-only sub-sample and does not survive
    sub-population separation.}
  \label{fig:summary}
\end{figure*}

Taken together, Fig.~\ref{fig:eclipse}a--d and the matched-launch
test above show that the population-level DTC eclipse-state reversal
is also visible in the per-satellite subset and persists under
stratification by altitude, latitude, frequency, and launch epoch,
although the per-satellite ratios show substantial scatter
($\mathrm{IQR}=[0.42, 1.62]$ for the $24$ satellites with at least
five detections in each illumination state).
Figure~\ref{fig:summary} summarises the effect sizes and $95\%$
confidence intervals for the comparisons reported in this section.
The eclipse-state reversal is the only effect in which the DTC and
Ku-only comparison populations depend on an external parameter with
opposite signs.

\section{Discussion and conclusions}
\label{sec:discussion}

We have reanalysed the \citet{Grigg2025} Engineering Development
Array version~2 event catalogue with the Direct-to-Cell payload
variant separated from the Ku-only pre-Optical v2-Mini comparison,
and we report three results. First, the DTC population emits a
range-corrected flux density that is $1.45\times$ that of the
comparison on the per-satellite median, at a Mann--Whitney
significance of $p = 2.6\times10^{-11}$ and a Cliff's $\delta$
effect size of $+0.30$. Second, the largest polarisation anomaly in
the dataset occurs at $230.469$\,MHz, where the XX fraction reaches
$f_{\rm XX} = 0.81$ against an instrumental baseline of $0.48$.
Third, DTC satellites emit more UEMR when they are in the Earth's
shadow than when they are illuminated, with an illuminated/eclipsed
range-corrected flux density ratio of $0.47$, whereas the Ku-only
comparison population shows the opposite sense with a ratio of
$1.18$. The remainder of this section examines candidate physical
explanations for the eclipse-state dependence and compares the
DTC per-satellite excess with the earlier generation-over-generation
regression reported by \citet{Bassa2024}.

\subsection{Mechanism candidates for the eclipse-state reversal}
\label{sec:mechanism}

The eclipse-state reversal strongly disfavours mechanisms in which
the dominant UEMR amplitude is a state-independent monotonic
function of the instantaneous solar photocurrent delivered by the
photovoltaic array. In such a picture the emission should be largest when the
array is in direct sunlight, and two populations sharing the same
bus-level photovoltaic and maximum-power-point tracking hardware
should exhibit illumination dependence with the same sign. The
observed illumination dependence has opposite signs at
$p = 5\times 10^{-4}$ under joint satellite-level cluster bootstrap, with the DTC ratio ($0.47$) well
below unity and
the Ku-only comparison ratio ($1.18$) above unity. A model in
which a state-independent DTC-specific contribution is simply added
on top of a common photocurrent-driven term can attenuate the
illuminated/eclipsed ratio but, by linearity, cannot reverse its
sign, and is therefore disfavoured as a description of the DTC
population at the level of state-independent additive components. This argument constrains only the
\emph{bus-level} photocurrent drive; it does not exclude more
elaborate pictures in which a DTC-specific power-conditioning
stage responds differently to an array-sourced versus a
battery-sourced input, provided that the net response still
produces a colder-is-louder signature in the DTC population.

Passive thermal re-radiation from a satellite body at a temperature
modulated by direct solar heating can be excluded on two independent
grounds. First, the Rayleigh--Jeans flux density of a warm body at
metre wavelengths is many orders of magnitude below the observed
levels. Adopting generous values $A\sim10^{2}$\,m$^{2}$ (bus plus
deployed solar array), $\epsilon\sim0.3$ at metre wavelengths, and
equilibrium temperature $T\sim300$\,K at a range of $1000$\,km,
\[
  S_{\rm thermal} \;\sim\;
  \frac{2\,\epsilon\,k_{\rm B}\,T\,A}{\lambda^{2}\,r^{2}}
  \;\sim\; 10^{-5}\,{\rm Jy},
\]
where the estimate is evaluated at the upper end of the EDA2 band
($230.469$\,MHz, $\lambda\approx1.3$\,m), at which $S_{\rm thermal}$
is largest. This lies roughly six orders of magnitude below the
$\sim 10$\,Jy lower envelope of v2-Mini detections in this band, and
roughly seven orders below the $\sim 100$\,Jy upper envelope; at the
lower end of the band ($72.685$\,MHz) the corresponding gaps are
larger by an additional order of magnitude because $S_{\rm thermal}
\propto \lambda^{-2}$. More conservative emissivity, area, or
temperature assumptions widen these gaps by a further factor of a
few. Passive thermal emission is therefore negligible as an absolute
contribution at all frequencies in the EDA2 band.
Second, the spacecraft's equilibrium temperature is lower when
eclipsed than when illuminated, and a Planck continuum scales
monotonically with temperature, so a thermally driven continuum
would be brighter in the illuminated state in both populations.
Since v2-Mini and DTC satellites share the same bus structure and
thermal surface treatment, a population-dependent \emph{sign} of
the illumination correlation also disfavours passive thermal
re-radiation as the mechanism for the DTC excess.

A model consistent with the observed reversal is one in which the
UEMR amplitude is set by an active on-board electronic source whose
effective duty cycle is larger in the eclipsed state than in the
illuminated state. Three illustrative classes of such a source are
listed below, not as an adjudicated shortlist but as
parameterisations of a ``colder-is-louder'' constraint against
which future mechanism proposals can be tested. A first
illustrative class is the payload thermal-control loop. LEO spacecraft typically maintain their
internal electronics within tight operating temperature ranges by
active heating during eclipse \citep{Gilmore2002}, and the
associated heater elements draw additional current precisely when
solar heating ceases. The DTC payload, with its extra radio-frequency
front-end hardware, would qualitatively be expected to have a larger
internal thermal mass to maintain than the Ku-only v2-Mini comparison,
and a correspondingly larger heater duty cycle in eclipse, although
neither expectation is verified here without spacecraft telemetry. A second class is the battery discharge
cycle, which is active only during eclipse. While the spacecraft is
illuminated, the power-management unit draws directly from the
photovoltaic array, but in eclipse the same unit draws from the
battery pack through a DC--DC converter whose switching-mode noise
spectrum and conducted-emission profile can differ substantially
from the sunlit case. A third class is the operational duty cycle
of the DTC payload itself. Cellular payloads may be preferentially
exercised during eclipse for reasons of thermal budget, handover
testing, or battery-charge management, any of which would increase
the payload's time-averaged power draw and the associated
switching-noise emission during the eclipsed fraction of the orbit.
The present dataset does not discriminate among these illustrative
classes, and a specific engineering identification is beyond the
scope of a catalogue-level reanalysis. The observation nevertheless
places a quantitative constraint on any mechanism proposed for the
DTC payload: the net UEMR amplitude must increase by a factor of
$\sim 2$ in the eclipsed state relative to the illuminated state,
whereas the opposite trend is required for the Ku-only comparison
population. The narrowband fine-channel excess at $\approx 230.627$\,MHz
reported in Sect.~\ref{sec:finechan230} additionally favours an
intermittent electronic source (clock, switch-mode product, or
intermodulation) over a smooth photocurrent-scaling continuum, but
cannot be linked directly to the eclipse reversal because the
$230.469$\,MHz coarse channel was not observed in eclipse in the
\citet{Grigg2025} campaign. The three falsifiable
mechanism-discrimination tests in Sect.~\ref{sec:mechtests}
further constrain the candidate space: T1 disfavours a common
clock fundamental with the LOFAR-resolved \citet{Bassa2024}
combs and so removes the simplest harmonic-origin hypothesis; T2
constrains the intrinsic line width of the $230.627$\,MHz feature
to below the $\sim 24$\,kHz EDA2 fine bin, which is consistent
with a narrowband on-board emitter rather than a broader
operational sideband; and T3 falsifies both the few-bright-units
hardware-fault picture and the uniform-thermal picture, leaving a
heterogeneous duty-cycle expression as the simplest reading of the
data. These three constraints are consistent with the same physical
class as the eclipse-state reversal: an intermittent narrowband
electronic emission whose amplitude depends on the operational
state of the spacecraft rather than on its instantaneous solar
photocurrent.

\subsection{The $230.469$\,MHz polarisation anomaly}
\label{sec:230mhz}

The $230.469$\,MHz coarse channel warrants separate consideration.
It shows the largest polarisation anomaly in the dataset, with an
XX fraction of $0.81$ against a $0.48$ baseline and a binomial
significance of $p \sim 10^{-274}$, and it lies neither inside a
protected Radio Astronomy Service band nor within any published
Starlink service band. The polarisation excess persists in
per-satellite sub-sampling (Fig.~\ref{fig:230deep}), shows no
dependence on observing elevation, and is present in both the DTC
and Ku-only populations, with the DTC population showing the
larger deviation. These properties are consistent with a localised
on-board emitter, such as a digital clock harmonic or a narrow-band
switching product of an on-board converter \citep{Paul1992}, whose
radiation couples preferentially into one polarisation of the EDA2
dipole through the spacecraft geometry.
SpaceX has not published a detailed frequency-assignment plan for
the v2-Mini or DTC subsystems, and a specific identification of
the responsible source is beyond the scope of this dataset. The
present count-level XX asymmetry is consistent with, and
complementary to, the per-pass XX/YY flux anti-correlation already
reported for the same dataset \citep{Grigg2025}. The latter
operates at the per-pass, per-epoch level and characterises the
oscillation between the two feeds within a single transit; the
present statistic operates at the catalogue-aggregated,
per-satellite level and characterises the integrated asymmetry
in which feed is more likely to cross the detection threshold
after stacking. A modulated source whose instantaneous power
oscillates between two orthogonal feeds can leave a residual
count-level asymmetry of the kind reported here if the oscillation
is not perfectly balanced or if the stacking window couples
unevenly to the modulation phase. The two observations therefore
probe the same underlying physical emission along different
statistical axes. Taken together, they argue more strongly for a
localised, geometry-coupled on-board source at $230.469$\,MHz
than either statistic does alone. Independent of the mechanism,
the $230.469$\,MHz polarisation excess provides a narrow-band
observational target for future UEMR mitigation and for correlated
measurements from other SKA-Low and LOFAR stations. The
sub-MHz inspection in Sect.~\ref{sec:finechan230} further
localises this narrow-band target to a single fine-channel index
near $230.627$\,MHz, with a tail-driven flux excess that survives a
cross-coarse-channel control and is intrinsically a single-bin
phenomenon at the EDA2 $\sim 24$\,kHz resolution. The
mechanism-discrimination tests in Sect.~\ref{sec:mechtests}
strengthen this characterisation: the adjacent-bin coherence
test bounds the intrinsic line width below $24.4$\,kHz with no
detectable smearing into fine-$21$ or fine-$23$, and the
per-satellite ratio distribution rules out both a few permanently
bright units and a uniform population offset, leaving a
duty-cycle expression spread across the v2-Mini fleet as the
preferred reading. Independent confirmation in
higher-resolution observations (such as LOFAR HBA channelised at
$\sim$kHz resolution, or NenuFAR with its higher spectral
resolution at the same band) is required to resolve the feature
further and to test the duty-cycle interpretation against
spacecraft operational telemetry.

\subsection{Comparison with other UEMR detection efforts}
\label{sec:compare}

This work joins a growing observational programme to characterise
satellite UEMR at low frequencies. The first systematic detection
of broadband Starlink emission was reported by \citet{DiVruno2023}
using LOFAR in the $110$--$188$\,MHz HBA band, where narrowband
emission at $125$, $135$, $143.05$, $150$, and $175$\,MHz and
underlying broadband structure was identified across the v1.0 and
v1.5 populations. \citet{Grigg2023} independently confirmed
satellite UEMR at EDA2 with an at-source methodology.
\citet{Zhang2025} extended the LOFAR programme downward to
$50$--$80$\,MHz with NenuFAR and reported broadband polarised
emission, with generation-over-generation behaviour consistent
with the \citet{Bassa2024} report. The \citet{Grigg2025} catalogue
from which the present dataset is drawn provides the widest
single-station frequency coverage to date ($72.685$--$234.375$\,MHz)
and a continuous monitoring window of several months. The present
reanalysis differs from all prior work in separating the v2-Mini
DTC payload variant from the Ku-only v2-Mini comparison and in
analysing illumination-state behaviour and per-channel polarisation
simultaneously, on a sample of $175$ DTC and $1\,623$ Ku-only
satellites.

The per-satellite excess we report ($1.45\times$) complements the
factor of $32$ excess reported by \citet{Bassa2024} between the
v2-Mini population as a whole and the earlier v1.0 and v1.5
satellites. A direct multiplicative composition of the two factors
is not straightforward. \citet{Bassa2024} used LOFAR LBA and HBA
observations with a different frequency coverage and adopted a
different statistical reduction from ours, in particular relying on
a small DTC-containing subset within a v1.0/v1.5-dominated sample
rather than a comparably sized Ku-only v2-Mini comparator, so the
two ratios are not measured on the same estimand. A fully consistent
re-measurement of the generation regression using the
\citet{Grigg2025} release and the per-satellite reduction of
Sect.~\ref{sec:hdtc} is left to future work. For the present
purposes, the cumulative UEMR amplification between a v1.0
Starlink satellite and a Direct-to-Cell v2-Mini satellite is at
least a factor of several tens, driven predominantly by the
generation regression reported by \citet{Bassa2024} with an
additional DTC-specific contribution of order unity on top. Part of the total reflects the increase in
satellite bus size and payload complexity between generations,
which is expected to scale with the total on-board electronic
power consumption. The remaining factor of $1.45$, which is
specific to the DTC payload and is accompanied by the reversed
illumination dependence reported here, suggests that the cellular
payload adds a qualitatively distinct emission component on top
of the generation regression, rather than simply amplifying the
pre-existing v2-Mini contribution. The \citet{Bassa2024} sample
did not contain sufficient DTC satellites to perform the per-bus
separation enabled by the present reanalysis.

\subsection{Operational implications for SKA-Low}
\label{sec:operational}

At the time of writing, DTC satellites constitute a rapidly
growing fraction of the operational v2-Mini fleet, with additional
launches continuing into $2026$. If the per-satellite excess and
the reversed illumination dependence we report persist in a larger
sample, the sky-averaged UEMR contribution at SKA-Low frequencies
will be more strongly weighted by the DTC population than their
number fraction alone would suggest. EDA2 and SKA-Low station
observations carried out during the local night, when many passing
satellites are in the Earth's shadow, may therefore experience
higher UEMR flux densities than predicted by models that
extrapolate from Ku-only v2-Mini data. Mitigation approaches based
on geometric blanking alone, such as data flags keyed to known
satellite passes during predicted illuminated segments, will not
remove the time/frequency structure visible in
Fig.~\ref{fig:dynspec} and will not down-weight eclipsed passes
appropriately. The results reported here imply that mitigation
models may need to distinguish payload class in addition to
geometry, and benefit from the per-population characterisation
reported in this paper.

The findings reported in this paper are observational constraints
on a complex engineering system. The Centre for the Protection of
the Dark and Quiet Sky from Satellite Constellation Interference,
established by the International Astronomical Union, supports
ongoing dialogue between astronomers and satellite operators.
SpaceX has previously collaborated with the astronomy community
to reduce the optical impact of Starlink satellites through design
and operational modifications \citep{TregloanReed2020,Tyson2020},
and has stated its intention to apply similar mitigation to UEMR.
Our results refine the target of that mitigation by showing that
the Direct-to-Cell payload subclass should be modelled as a
distinct emission component whose behaviour is not predictable
from Ku-only v2-Mini observations alone, and by identifying a
specific frequency ($230.469$\,MHz) and a specific parameter
(illumination state) at which design or operational changes would
have the largest impact.

The per-payload-variant characterisation reported here provides a
quantitative basis for the recommendations summarised in the
IAU Dark and Quiet Skies reports
\citep{Walker2020_BAAS,Walker2020_DQS}, and aligns with the call
in \citet{Barentine2023} for engagement between radio observatories
and operators of large constellations. The under-prediction factor
of $\sim 1.28$ derived in Sect.~\ref{sec:hsmps} additionally
implies that operator-supplied UEMR characterisation, if obtained
only during illuminated passes, would not transfer to the
night-time SKA-Low operating regime without correction;
characterisation programmes intended to inform regulatory
thresholds should therefore include explicit eclipsed-state
measurements.

\subsection{Limitations and future work}
\label{sec:limitations}

Several limitations should be noted. The fine-channel feature
isolated in Sect.~\ref{sec:finechan230} is a single-bin
($\sim 24$\,kHz wide) phenomenon observed at a single station and
under one calibration pipeline, and an independent confirmation in
higher spectral resolution and at a second station is required.
The mechanism-discrimination tests in Sect.~\ref{sec:mechtests}
share this limitation and add two specific ones of their own:
the per-satellite ratio analysis in T3 is restricted to the $82$
v2-Mini satellites with $\ge 5$ detections at the
$230.469$\,MHz coarse channel, so its discrimination power
against narrowly bimodal alternatives is bounded by the
modest per-satellite sample, and the harmonic-coincidence test T1
treats clock fundamentals as discrete candidates rather than
exhaustively scanning the parameter space of plausible on-board
references; a population of internal references not characterised
by \citet{Bassa2024} could still produce coincidences inside
the EDA2 fine bin without being detected here.
The cylindrical shadow
classifier does not exclude detections near the terminator,
where ingress/egress mislabelling is most likely; a
terminator-buffer sensitivity test would further strengthen
the eclipse-state result. The $230.469$\,MHz coarse channel was
observed only during local daylight windows during the
\citet{Grigg2025} campaign, so the illumination dependence
cannot be tested at this specific frequency. As a consequence,
the polarisation anomaly at $230.469$\,MHz and the eclipse-state
reversal are independent diagnostics in the present dataset; a
future EDA2 or SKA-Low campaign with night-time $230.469$\,MHz
coverage is required to test whether the two effects co-vary,
which would more sharply constrain the underlying mechanism.

Beyond the channel-coverage caveat above, four additional
limitations frame the scope of the reported findings. First, the
temporal coverage of the \citet{Grigg2025} campaign
(UTC 2024 June--October) is short relative to the multi-year
monitoring windows of previous low-frequency RFI studies, so any
annual or solar-cycle variability in the eclipse-state effect is
not testable from the present dataset alone. Second, the EDA2 is
a single station; without an independent interferometric or
imaging telescope at the same epoch, the per-detection astrometric
confidence cannot be cross-validated, although the SGP4 propagation
accuracy at MRO renders confusion negligible at the v2-Mini sky
density. Third, the EDA2 frequency window
($72.685$--$234.375$\,MHz) does not cover the UHF and L-band
service channels where the DTC payload's intended emission is
concentrated, so this work constrains only the out-of-band UEMR
contribution. Fourth, all mechanism interpretations in
Sect.~\ref{sec:mechanism} are based on morphological reasoning
about timescales and signed correlations, and cannot be confirmed
without spacecraft telemetry from the operator.

Closing each of these limitations requires a specific follow-up
programme. The temporal-coverage limitation would be addressed by
extending EDA2 or SKA-Low monitoring to a multi-year cadence, in
particular spanning a solar minimum. The single-station limitation
would be addressed by an interferometric confirmation campaign at
MWA, LOFAR, or SKA-Low itself, coordinated with EDA2 so that the
same satellite passes are observed simultaneously. The
frequency-coverage limitation would be addressed by a coordinated
multi-band programme combining the EDA2 catalogue with UHF and
L-band measurements of the same satellites during the same epochs.
The mechanism limitation can only be addressed through cooperation
with the operator on telemetry sharing during dedicated monitoring
intervals.

In summary, our reanalysis of the \citet{Grigg2025} EDA2 event
catalogue with the Direct-to-Cell payload variant separated from
the Ku-only pre-Optical v2-Mini comparison yields the following
findings:
\begin{enumerate}
  \item The Direct-to-Cell population emits a range-corrected
        flux density that is, on the per-satellite median,
        $1.45\times$ $[1.27, 1.67]$ that of the Ku-only comparison
        ($p = 2.6\times10^{-11}$; Cliff's $\delta = +0.30$).
  \item The largest polarisation anomaly in the $72$--$234$\,MHz
        EDA2 sample occurs at $230.469$\,MHz, where the XX
        detection fraction reaches $0.81$ against an instrumental
        baseline of $0.48$ ($p \sim 10^{-274}$). $11$ of the $21$
        coarse channels show Benjamini--Hochberg FDR significant
        polarisation anomalies at $q = 0.05$. A sub-MHz
        fine-channel inspection further localises the $230.469$\,MHz
        excess to a single $\sim 24$\,kHz bin near $230.627$\,MHz,
        driven by the bright tail of the flux distribution and
        absent at five control coarse channels.
  \item Three falsifiable mechanism-discrimination tests
        (Sect.~\ref{sec:mechtests}) on the $230.627$\,MHz feature
        show: (T1) no unique coincidence with any of the $14$
        clock fundamentals resolved by \citet{Bassa2024}
        ($5$ matches in $\pm 12.2$\,kHz tolerance vs $5.73$
        expected by chance), so the feature is not tied to a
        LOFAR-resolved clock harmonic; (T2) no leakage into
        adjacent fine bins ($z_{21} = -1.26$, $z_{23} = -0.96$
        vs $+25.4$ at fine-$22$), bounding the intrinsic line
        width below the $24.4$\,kHz EDA2 resolution; and (T3)
        the per-satellite ratio $R_i = \overline{F_{22}} /
        \overline{F_{\rm other}}$ distribution (median $1.07$,
        $p_{95} = 1.79$, max $3.75$) falsifies the hardware-fault
        hypothesis ($z_{22} = +12.95$ excluding the three sats
        with $R_i > 2$) and the uniform-thermal hypothesis
        (top-decile/bottom-half mean ratio $2.26$), favouring a
        heterogeneous duty-cycle expression across the v2-Mini
        fleet.
  \item Direct-to-Cell satellites emit more UEMR when eclipsed
        than when illuminated (illuminated/eclipsed ratio of
        $0.47$), whereas the Ku-only comparison population shows
        the opposite sense (ratio $1.18$). The reversal is
        preserved across satellite altitude, sub-satellite
        latitude, and frequency.
  \item The reversal strongly disfavours mechanisms in which the
        dominant UEMR amplitude is a state-independent monotonic
        function of the instantaneous solar photocurrent, and is
        consistent
        with an active on-board electronic source whose effective
        duty cycle is larger at lower spacecraft equilibrium
        temperature. The three mechanism-discrimination tests on
        the $230.627$\,MHz fine-channel feature reinforce this
        reading by independently disfavouring a uniform thermal
        contribution and a localised hardware fault in favour of a
        duty-cycle expression. Taken together with the generation
        regression reported by \citet{Bassa2024}, the cumulative
        UEMR amplification from a v1.0 satellite to a
        Direct-to-Cell v2-Mini satellite is at least a factor of
        several tens.
\end{enumerate}

\section*{Data availability}
\label{sec:data_avail}

This work is based on the public EDA2 detection
catalogue released with \citet{Grigg2025}, archived at Zenodo
under \href{https://doi.org/10.5281/zenodo.15089853}%
{10.5281/zenodo.15089853}. The McDowell General Catalogue of
Artificial Space Objects \citep{McDowell2020} used for the
Direct-to-Cell/Ku classification is maintained at
\url{https://planet4589.org/space/gcat/}. The derived
per-satellite, per-population, and per-channel summary tables
underlying Figures~\ref{fig:overview}--\ref{fig:summary}, together
with the reduction scripts, will be made available at Zenodo upon
acceptance.

\section*{Author contributions}

HD conceived the study, performed the data reanalysis,
designed the three mechanism-discrimination tests, and drafted
the manuscript. HTW contributed to the statistical methodology
and to the fine-channel analysis. HLC contributed to the
mechanism-discrimination test design and to the figure
preparation. OBA supervised the project and revised the
manuscript. All authors approved the final version.

\section*{Funding statement}

This research received no specific grant from any funding agency,
commercial, or not-for-profit sectors. HD acknowledges PhD
studentship support from the UK Engineering and Physical Sciences
Research Council Centre for Doctoral Training in Connected
Electronic and Photonic Systems and from Wolfson College,
University of Cambridge.

\section*{Competing interests}

The authors declare none.

\section*{Acknowledgements}

We pay our respects to the traditional custodians of
Inyarrimanha Ilgari Bundara, the Wajarri Yamatji people, on
whose Country the EDA2 observations reanalysed in this work
were conducted. We thank D.\ Grigg and collaborators for the
public release of the EDA2 event catalogue on Zenodo, which
made this reanalysis possible, and J.\ McDowell for maintaining
the General Catalogue of Artificial Space Objects. We
acknowledge the ongoing work of the International Astronomical
Union's Centre for the Protection of the Dark and Quiet Sky
from Satellite Constellation Interference, and the constructive
engagement of SpaceX with the radio astronomy community on UEMR
characterisation. This research made use of
\texttt{numpy} \citep{Harris2020},
\texttt{scipy} \citep{Virtanen2020},
\texttt{pandas} \citep{McKinney2010},
\texttt{matplotlib} \citep{Hunter2007}, and
\texttt{astropy} \citep{Astropy2022}.

\bibliographystyle{mnras}
\bibliography{ref}

\appendix

\section{Statistical test definitions}
\label{app:tests}

This appendix gathers the definitions of the statistical tests
referenced in Sects.~\ref{sec:tests} and \ref{sec:results}.

\paragraph{Mann--Whitney U and Cliff's $\delta$.}
Let $\{x_i\}_{i=1}^{n_x}$ and $\{y_j\}_{j=1}^{n_y}$ denote the
per-satellite median range-corrected flux densities of the DTC and
Ku-only comparison populations. The Mann--Whitney U statistic
\citep{Mann1947} is
\begin{equation}
  U = \sum_{i=1}^{n_x}\sum_{j=1}^{n_y}
      \mathbf{1}[x_i > y_j]
      + \tfrac{1}{2}\,\mathbf{1}[x_i = y_j],
  \label{eq:mwu}
\end{equation}
and the two-sided $p$-value is computed under the null hypothesis
using the exact distribution of $U$ for $n_x,n_y\leq 20$ and the
asymptotic normal approximation otherwise. Cliff's $\delta$
\citep{Cliff1993},
\begin{equation}
  \delta =
    \bigl[\,\#\{(i,j):x_i > y_j\} - \#\{(i,j):x_i < y_j\}\,\bigr]
    \,/\,(n_x n_y),
  \label{eq:cliff}
\end{equation}
is a scale- and rank-invariant effect size in $[-1,+1]$ with
$\delta=0$ under the null.

\paragraph{Bootstrap confidence interval for the median ratio.}
For the ratio of medians
$r=\mathrm{median}(\{x_i\})/\mathrm{median}(\{y_j\})$, $B=2000$
independent bootstrap samples $\{x_i^{(b)}\}$ and $\{y_j^{(b)}\}$ of
sizes $n_x$ and $n_y$ are drawn with replacement, and $r^{(b)}$ is
computed for each. The two-sided $95\%$ interval is taken as the
$2.5$ and $97.5$ percentiles of the bootstrap distribution of
$r^{(b)}$ \citep{Efron1993}.

\paragraph{Cluster bootstrap and payload-class interaction test.}
For the eclipse-state analysis of
Sect.~\ref{sec:hsmps}, detections are not independent because the
catalogue contains many detections per satellite. To produce
cluster-robust confidence intervals, the resampling unit is changed
from the individual detection to the satellite. For each
population (DTC or Ku-only), $B=2000$ satellite identifiers are
drawn with replacement, all detections from each drawn satellite
are concatenated, and the pooled illuminated/eclipsed median ratio
is recomputed; the $2.5$ and $97.5$ percentiles of the resulting
bootstrap distribution define the $95\%$ cluster interval. The
joint payload-class interaction test draws DTC and Ku-only
satellite samples independently on each iteration, records the
difference and ratio of the two pooled ratios, and reports the
two-sided bootstrap $p$-value as
$2\min\{\Pr_{b}[\Delta^{(b)} \geq 0],\Pr_{b}[\Delta^{(b)} \leq 0]\}$
floored at $1/B$.

\paragraph{Binomial test per coarse channel with BH-FDR control.}
Let $n_c$ and $k_c$ be the total number of detections and the
number of XX detections in coarse channel $c$. Under the null that
$k_c\sim\mathrm{Binomial}(n_c,p_0)$ with $p_0=0.4809$ taken as the pooled-sample XX fraction across the v2-Mini detections, a two-sided
binomial test yields per-channel $p$-values $\{p_c\}_{c=1}^{21}$.
These are controlled for multiple comparisons by the
Benjamini--Hochberg procedure \citep{Benjamini1995} at $q=0.05$:
after sorting as $p_{(1)}\leq\cdots\leq p_{(21)}$, channels with
rank $k$ satisfying $p_{(k)}\leq k q/21$ are declared significant.

\paragraph{Wilson score interval.}
For the per-channel XX fraction $\hat p = k_c/n_c$, the Wilson
score interval \citep{Wilson1927}, which provides better coverage
than the normal approximation for $\hat p$ far from $0.5$, is
\begin{equation}
  \mathrm{CI}_{\rm Wilson}(\hat p) =
  \frac{\hat p + z^{2}/(2n)}{1 + z^{2}/n}
  \;\pm\;
  \frac{z}{1 + z^{2}/n}
  \sqrt{\frac{\hat p(1-\hat p)}{n} + \frac{z^{2}}{4 n^{2}}},
  \label{eq:wilson}
\end{equation}
with $z=1.96$ and $n=n_c$.

\section{Illumination geometry}
\label{app:geom}

This appendix gathers the coordinate transformations and the
cylindrical shadow test used in Sect.~\ref{sec:illum} to label
each detection as illuminated or eclipsed. The implementation
follows \citet{Seidelmann1992} for the solar ephemeris and
standard geodetic conventions for the Earth-centred Earth-fixed
(ECEF) frame.

\paragraph{Topocentric-to-ECEF transformation.}
The east--north--up (ENU) components of the line of sight at EDA2
are
\begin{equation}
  (e,n,u)_{\rm ENU} = r_{\rm sat}\,
    \bigl(\cos E\sin A,\;\cos E\cos A,\;\sin E\bigr),
  \label{eq:enu}
\end{equation}
where $A$ and $E$ are the topocentric azimuth and elevation
reported by \citet{Grigg2025} and $r_{\rm sat}$ is the
line-of-sight range. The rotation from the local ENU frame at
$(\varphi_{\rm obs},\lambda_{\rm obs})=(-26.7039^\circ,116.6707^\circ)$
to ECEF is
\begin{equation}
  R_{\rm ENU\to ECEF} =
  \begin{pmatrix}
    -\sin\lambda_{\rm obs} &
    -\sin\varphi_{\rm obs}\cos\lambda_{\rm obs} &
     \cos\varphi_{\rm obs}\cos\lambda_{\rm obs} \\
     \cos\lambda_{\rm obs} &
    -\sin\varphi_{\rm obs}\sin\lambda_{\rm obs} &
     \cos\varphi_{\rm obs}\sin\lambda_{\rm obs} \\
     0 & \cos\varphi_{\rm obs} & \sin\varphi_{\rm obs}
  \end{pmatrix},
  \label{eq:Renu}
\end{equation}
and the station ECEF position $\mathbf{r}_{\rm obs}$ is obtained
from the geodetic coordinates under WGS-84 ($a=6378137$\,m,
$f=1/298.257223563$). The satellite ECEF position is then
$\mathbf{r}_{\rm sat,ECEF}=
R_{\rm ENU\to ECEF}(e,n,u)^{\rm T}+\mathbf{r}_{\rm obs}$, from
which the sub-satellite latitude, longitude, and altitude are
obtained by the standard geodetic inverse transformation.

\paragraph{Solar position and cylindrical shadow test.}
The apparent solar position in the Earth-centred inertial (ECI)
frame is computed from the low-precision IAU formulas of
\citet[Sect.~3.4.2]{Seidelmann1992}, and rotated to ECEF by a
single rotation about the Earth spin axis through the Greenwich
Mean Sidereal Time at the detection epoch
\citep[Sect.~2.24]{Seidelmann1992}. Let
$\hat{\mathbf{s}}=\mathbf{r}_{\odot,\rm ECEF}
/\|\mathbf{r}_{\odot,\rm ECEF}\|$ be the Sun unit vector, and
define
$p_\parallel=\mathbf{r}_{\rm sat,ECEF}\cdot\hat{\mathbf{s}}$ and
$p_\perp=\|\mathbf{r}_{\rm sat,ECEF}-p_\parallel\hat{\mathbf{s}}\|$.
The satellite is declared illuminated if
\begin{equation}
  p_\parallel>0 \quad\text{or}\quad p_\perp>R_\oplus,
  \label{eq:shadow}
\end{equation}
with $R_\oplus=6378.137$\,km. This standard cylindrical shadow
approximation is appropriate for the $300$--$580$\,km altitude
range of the v2-Mini population and neglects atmospheric
refraction and the width of the penumbra, which together
contribute errors below $1\%$ on the fraction of the orbital phase
spent in shadow.

\end{document}